\documentclass[12pt]{article}
\pdfoutput=1

\usepackage[utf8]{inputenc}
\usepackage{geometry}
\usepackage[colorlinks]{hyperref}
\usepackage{cite}
\hypersetup{pageanchor=true,citecolor=blue,urlcolor=blue,colorlinks=true,linkcolor=blue}
\usepackage[toc,page]{appendix}
\usepackage[]{graphicx}
\usepackage{amsmath}
\usepackage{mathrsfs}
\usepackage{amssymb}
\usepackage{array}
\usepackage{chngcntr}

\counterwithin*{equation}{section}

\usepackage{cleveref}

\newcommand{\be}{\begin{equation}}
\newcommand{\ee}{\end{equation}}
\newcommand{\bl}{\begin{align}}
\newcommand{\el}{\end{align}}
\newcommand{\bseq}{\begin{subequations}}
\newcommand{\eseq}{\end{subequations}}

\renewcommand{\l}{\lambda}
\renewcommand{\O}{\Omega}
\renewcommand{\o}{\omega}
\renewcommand{\k}{\kappa}
\renewcommand{\b}{\beta}
\renewcommand{\H}{\mathcal{H}}
\newcommand{\B}{\mathscr{B}}
\newcommand{\D}{\mathscr{D}}
\newcommand{\g}{\gamma}
\newcommand{\vk}{\varkappa}
\newcommand{\G}{\mathcal{G}}

\newcommand{\vf}{\varphi}
\newcommand{\gc}{\mathrm{g}}
\newcommand{\diff}{\mathrm{d}}
\newcommand{\e}{{\rm e}}
\newcommand{\bb}{\vf_{\mathrm{b}}}
\newcommand{\C}{\mathscr{C}}
\renewcommand{\th}{\mathop{\rm th}\nolimits}

\newcommand{\ch}{\mathop{\rm ch}\nolimits}
\newcommand{\sh}{\mathop{\rm sh}\nolimits}
\newcommand{\arctg}{\mathop{\rm arctg}\nolimits}

\begin{document}
\begin{titlepage}

\title{\vspace{-2cm}
\begin{flushright}
{\small FTPI-MINN-21-20,~~UMN-TH-4103/2,~~INR-TH-2021-021}
\end{flushright}
\vspace{1.5cm}
{\bf Black hole induced false vacuum decay:\\ 
The role of greybody factors}}

\author{Andrey Shkerin$^{a}$\footnote{ashkerin@umn.edu}~, 
Sergey Sibiryakov$^{b,c,d}$\footnote{ssibiryakov@perimeterinstitute.ca}\\[2mm]
{\small\it $^a$William I. Fine Theoretical Physics Institute, School
  of Physics and Astronomy,} \\ 
{\small\it University of Minnesota, Minneapolis, MN 55455, USA }\\
{\small\it $^b$Department of Physics \& Astronomy, McMaster
University,}\\
{\small\it Hamilton, Ontario, L8S 4M1, Canada}\\
{\small\it $^c$Perimeter Institute for Theoretical Physics, Waterloo,
 Ontario, N2L 2Y5, Canada}\\
{\small \it $^d$Institute for Nuclear Research of the Russian Academy
  of Sciences,}\\ 
{\small \it 60th October Anniversary Prospect, 7a, 117312 Moscow, Russia}
}

\date{}
\maketitle

\begin{abstract}
    We study false vacuum decay catalyzed by black holes. We consider
    a toy two-dimensional model of a scalar field with an unstable
    potential in the background of a dilaton black hole. A realistic
    black hole in four dimensions possesses the potential barrier for
    linear field perturbations. We model this barrier --- the greybody
    factor --- for spherically-symmetric perturbations in the toy
    model by adding a coupling between the scalar field and
    dilaton. We compute analytically the decay rate 
for the black hole in thermal equilibrium
    (Hartle--Hawking state) and for the radiating black hole in empty
    space (Unruh state). Our results show that, contrary to the
    Hartle--Hawking vacuum, the decay probability of the Unruh vacuum
    remains exponentially suppressed at all black hole
    temperatures. We argue that this result holds also in four
    dimensions. 
\end{abstract}

\thispagestyle{empty}
\end{titlepage}

\newpage 
\tableofcontents

\section{Introduction}

The problem of catalysis of vacuum decay by black holes \cite{Hiscock:1987hn,Berezin:1987ea,Arnold:1989cq,Berezin:1990qs} has recently received significant attention in view of its possible relevance for phenomenology \cite{Gregory:2013hja,Burda:2015isa,Burda:2015yfa,Burda:2016mou,Tetradis:2016vqb,Canko:2017ebb,Gorbunov:2017fhq,Mukaida:2017bgd,Kohri:2017ybt,Dai:2019eei,Hayashi:2020ocn,Miyachi:2021bwd}. In the Standard Model, the loop-corrected Higgs field potential may develop large negative values at large field values, which makes the low-energy electroweak vacuum metastable \cite{Flores:1982rv,Sher:1988mj,Isidori:2001bm,
1205.2893,1205.6497,1307.3536,Bednyakov:2015sca}. Requirement that the
lifetime of the vacuum exceeds the age of the Universe puts
constraints on the parameters of the Standard Model, of its possible
extensions, and of systems and environments which catalyze vacuum
decay, including black holes; see, e.g., \cite{Markkanen:2018pdo} for
a review. 

The catalyzing effect of a black hole (BH) is two-fold. First, it is a
local spacetime inhomogeneity. Hence, as many types of impurities, it
can facilitate nucleation of bubbles of true vacuum in its
vicinity. Second, BHs excite the quantum vacuum producing 
Hawking radiation. As any field excitation, this
radiation is expected to increase the decay rate. For small enough
BHs, the catalyzing effects due to curved geometry and quantum
excitations 
may be equally important.  

In the semiclassical regime, the vacuum decay is described by a
(complex) classical solution of field equations. The solution ---
bounce --- saturates the amplitude of transition from the false to the
true vacuum regions. It is important to note that the vacuum is
defined not only by classical field expectation values but also by the
state of quantum fluctuations around these values. Different false
vacuum states reveal themselves through different boundary conditions
imposed on the bounce. 

It is well-known how to obtain the bounce solution in equilibrium
systems
\cite{Coleman:1977py,Callan:1977pt,Coleman:1978ae,Coleman:1980aw}. One
rotates the system to the Euclidean time and looks for a regular
solution satisfying appropriate boundary conditions. In the case of
BHs, this prescription leads to a periodic Euclidean bounce whose
period is inversely proportional to the BH temperature $T_{\rm BH}$,
or to a static solution --- sphaleron. The solution with the smallest
Euclidean action dominates and describes vacuum
transitions catalyzed by a BH in the presence of thermal bath of
temperature $T_{\rm BH}$. This state of a BH in thermal equilibrium is
known as the Hartle--Hawking vacuum \cite{Hartle:1976tp}. The decay
rate of the Hartle--Hawking vacuum is not exponentially suppressed at
high temperatures, reflecting the fact that thermal field
fluctuations trigger vacuum decay with order-one probability once
their average energy exceeds the height of the barrier separating the vacua. 

Realistic BHs, however, are not in thermal equilibrium with their
environment. In particular, this is true for hypothetical small
primordial BHs that could exist in the early Universe but are
completely evaporated by now
\cite{GarciaBellido:1996qt,Fujita:2014hha,Dong:2015yjs,Allahverdi:2017sks, 
Lennon:2017tqq,Morrison:2018xla,Hooper:2019gtx,Carr:2020gox,Hooper:2020evu}. 
At 
the late stage of evaporation, a BH radiates at energies 
comparable to the Planck scale, much above the temperature possibly
attained in the primordial plasma. From the perspective of Standard
Model vacuum decay, such near-Planckian BHs are of principal
interest. This is because the bounce mediating decay of the
electroweak vacuum probes field values near the minimum of the Higgs
quartic coupling 
which, for the measured values of the Standard Model
parameters \cite{ParticleDataGroup:2020ssz}, approaches the Planck
scale.  

The above considerations motivate to look for an approach to vacuum
decay that can handle such non-equilibrium systems as a hot isolated
BH placed in a comparatively low energy environment. This system is
described by the Unruh state \cite{Unruh:1976db}. In the Unruh vacuum,
the BH emits Hawking radiation but does not receive anything from
asymptotic infinity. The Euclidean prescription described above is not
suitable for a BH in the Unruh vacuum. 

It is clear why the decay of the Hartle--Hawking vacuum is unsuppressed
at high temperatures. It is much less clear if the Unruh
vacuum decay is unsuppressed at high temperatures. From studying
vacuum transitions in thermal equilibrium we know that to trigger decay one
needs to form a field fluctuation that is coherent on a certain length
scale typically associated with the Compton wavelength of the free
field. Such fluctuation is represented by the sphaleron. It is not
clear what the analog of the sphaleron in the Unruh vacuum is.  

In Ref.~\cite{Shkerin:2021zbf} we have suggested a
method to compute the Unruh vacuum decay rate using complex tunneling
solutions
\cite{Miller,Rubakov:1992ec,Bonini:1999kj,Bezrukov:2003tg,Bramberger:2016yog}.\footnote{See
  \cite{Turok:2013dfa,Cherman:2014sba,Plascencia:2015pga,Andreassen:2016cff,Andreassen:2016cvx}
  for related approaches to tunneling.} The method accounts for the
quantum state of the false vacuum and, hence, it allows one to
discriminate between the Hartle--Hawking and Unruh vacua. The key
ingredient is the correspondence between the quantum state and the
boundary conditions on the bounce. The boundary conditions
turn out to be the same as for the time-ordered Green's function in
the corresponding vacuum. 
In the case of Hartle--Hawking vacuum the method reproduces the Euclidean time formalism used in previous works on BH catalysis of vacuum decay \cite{Gregory:2013hja,Burda:2015isa,Burda:2015yfa,Burda:2016mou}.
The Unruh vacuum provides a genuinely new application of the method.
For explicit calculations
Ref.~\cite{Shkerin:2021zbf} adopted a toy
two-dimensional model consisting of a real
scalar field in the background of a dilaton BH.\footnote{The
  back-reaction of the tunneling field on spacetime geometry is taken
  to be negligible in \cite{Shkerin:2021zbf}.} Thanks to the choice of the
tunneling potential, bounce solutions and associated decay rates were
found analytically, both for the Hartle--Hawking and Unruh vacua,
and both in the BH vicinity and far from it. 

In the model of \cite{Shkerin:2021zbf}, the exponential suppression of
the Unruh vacuum decay rate vanishes at high temperatures. We would
like to see if this result holds also in a realistic case like the
Schwarzschild BH in four dimensions. 
The model of \cite{Shkerin:2021zbf} lacks two important features of
the realistic setup. First, a field in Schwarzschild background feels
a centrifugal barrier whose height is inversely proportional to the
square of the BH size and, hence, grows with the BH temperature.
There is no such barrier in two dimensions. Its presence
in four dimensions may significantly affect 
the Unruh vacuum decay rate at high
temperatures~\cite{Gorbunov:2017fhq}.
Second, the flux of Hawking quanta in four dimensions spreads
inversely proportional to the area of the sphere encompassing the BH
at a given distance. This
reduces the density of Hawking radiation away from the BH and is expected
to further suppress the tunneling rate of Unruh
vacuum~\cite{Gorbunov:2017fhq,Johnson_priv}. 

The present paper is a follow-up of Ref.~\cite{Shkerin:2021zbf}. Its goal is to compute the rate of decay of the
Unruh vacuum in a model containing the temperature-dependent barrier
for massive scalar modes. This will bring us one step closer to the
problem of vacuum decay in the realistic BH background in four
dimensions. To emulate the barrier, we modify the model of
\cite{Shkerin:2021zbf} by adding a coupling between the tunneling
scalar field and the dilaton. Note that the area growth in four
dimensions remains unaccounted for in our setup. We will discuss
qualitatively its possible effect on vacuum decay at the end of the
paper. 

The scalar-dilaton coupling constant that controls the strength of the
barrier is a free parameter of the model. When it is zero, the model
reduces to the one studied in \cite{Shkerin:2021zbf}. In particular,
the exponential suppression of the Unruh vacuum goes to zero at high
BH temperatures. The main result of this paper is that, whenever the
coupling is non-zero, the decay of the Unruh vacuum remains
exponentially suppressed at all temperatures.\footnote{More precisely,
  the lower bound on the coupling comes from the requirement for the
  tunneling action to be large at all $T_{\rm BH}$. We will see that
  the bound is proportional to another coupling constant that controls the
  semiclassical expansion and can be made arbitrarily small.} 

The paper is organized as follows. In sec.~\ref{sec:setup} we
recapitulate the results of \cite{Shkerin:2021zbf}, 
and describe the toy model used to study
vacuum decay. We outline the technique to find the tunneling
solution and illustrate it in the case of vacuum decay in flat
space. In sec.~\ref{sec:HH} we study the decay of the Hartle--Hawking
vacuum in the BH vicinity. The results of this study are important in
two respects. First, it is instructive to compare transitions from the
Hartle--Hawking vacuum to those from the Unruh vacuum, since the
difference between the two states is entirely due to the different
population of Hawking quanta, i.e., due to the different quantum
vacuum structure.
Second, and more importantly, the Hartle--Hawking vacuum decay can be straightforwardly studied in four dimensions.
Hence, we can directly compare the physics of
tunneling in our toy model for the different values of the
scalar-dilaton coupling and in a four-dimensional scalar field theory
in the Schwarzschild background. We can then select the values of the
coupling parameter for which we see the best agreement in the behavior
of the two systems. 

In sec.~\ref{sec:U} we study the decay of the Unruh vacuum. We
analytically construct the Unruh bounces for the range of the
scalar-dilaton couplings emulating the four-dimensional behavior
and compute the associated decay rates. Not
all values of BH temperature admit analytic bounce
solution. When no such solution is available, we employ a 
stochastic estimate of the decay rate. We find that the exponential
suppression of 
the Unruh vacuum decay rate is 
constant in the high temperature limit.  

We conclude in sec.~\ref{sec:disc}. The main text is accompanied by several
appendices.

\section{Setup}
\label{sec:setup}

\subsection{Background geometry}
\label{ssec:geometry}

We consider a real scalar field in the background of a dilaton BH in
two dimensions \cite{Callan:1992rs}. 
The dilaton BH is characterized by the temperature
$T_{\rm BH}$ and mass $M$ which are a priori independent of each
other.\footnote{This is different from the four-dimensional Schwarzschild
case where $T_{\rm BH}$ and $M$ are related by 
$$T_{\rm BH}=\frac{M_{\rm
  Pl}^2}{8\pi M}$$ with $M_{\rm Pl}$ the Planck mass.} 
The BH background is set up by the metric $g_{\mu\nu}$ and the
dilaton field $\phi$. As discussed in \cite{Shkerin:2021zbf}, the only
region that is relevant for vacuum decay is the 
patch of the BH spacetime outside the horizon. 
It is convenient to introduce tortoise
coordinates $(t,x)$ covering this patch. Then the metric is read off
from the line element 
\begin{equation}
\label{ds}
    \diff s^2=\O(x)(-\diff t^2+\diff x^2) \;,
\end{equation}
where the conformal factor takes the form
\begin{equation}
\label{O}
    \O(x)=\frac{1}{1+\e^{-2\l x}} \;.
\end{equation}
The parameter $\l$ is related to the BH temperature as $\l=2\pi T_{\rm
  BH}$. For the sake of brevity, we will refer to $\l$ itself as
temperature. The horizon is located at $x\to-\infty$. Near horizon,
the conformal factor behaves as $\O(x)\approx\e^{2\l x}$. Substituting
this to \cref{ds}, we obtain the two-dimensional Rindler metric. Thus, the
near-horizon region is approximated by the Rindler spacetime. 
The physical size of this region is 
\begin{equation}
\label{l_h}    l_h\sim\int_{-\infty}^0\diff x\sqrt{\O}\sim\frac{1}{\l} \;.
\end{equation}
In the opposite limit, $x\to\infty$, the metric is asymptotically flat. Finally, the dilaton profile is given by
\begin{equation}
\label{phi}
    \phi=-\frac{1}{2}\ln\left[ \frac{M}{2\l}\left(1+\e^{2\l x} \right) \right] \;.
\end{equation}
In appendix \ref{app:dilaton} we discuss dilaton BHs in more details. The vacuum boundary conditions are imposed in the remote past, $t\to-\infty$, on both sides of the physical patch, $x\to\pm\infty$.

\subsection{Massive scalar field with a dilaton coupling}
\label{ssec:linear}

To study tunneling in the BH background, we consider the following scalar field theory
\begin{equation}
\label{action_gen}
    S=\frac{1}{\gc^2}\int\diff^2x\sqrt{-g}\left( -\frac{1}{2}g^{\mu\nu}\partial_\mu\vf\partial_\nu\vf-\frac{m^2\vf^2}{2}-Q\e^{2\phi}\vf^2-V_{\rm int}(\vf) \right) \;.
\end{equation}
Here $V_{\rm int}(\vf)$ is the interaction part of the tunneling
potential to be specified below. So far it suffices to assume that the
false vacuum is located at $\vf=0$ where the potential vanishes,
$V_{\rm int}(0)=0$. Next, $Q>0$ is the nonminimal coupling (of mass
dimension 2) of the scalar field to the dilaton. We will see shortly
that this coupling gives rise to the temperature-dependent barrier in
the potential for linearized field perturbations, which is similar to
the centrifugal barrier in the four-dimensional
Schwarzschild spacetime. We will refer to it as ``dilaton barrier'' in what follows. The small
coupling constant $\gc\ll 1$ controls the semiclassical expansion in
the model.  

In the background specified by \cref{ds,O,phi}, the action (\ref{action_gen}) becomes
\begin{equation}
\label{action_O}
    S=\frac{1}{\gc^2}\int\diff t\diff x\left( -\frac{1}{2}\eta^{\mu\nu}\partial_\mu\vf\partial_\nu\vf-\frac{1}{2}\O m^2\vf^2 - \frac{Q\O'}{M}\vf^2 -\O V_{\rm int}(\vf) \right) \;,
\end{equation}
where $\eta^{\mu\nu}=\text{diag}(-1,1)$ is the Minkowski metric. We see that the dependence on the background is contained entirely in the (position-dependent) potential of the field $\vf$.

\begin{figure}[t]
    \centering
    \includegraphics[width=0.6\linewidth]{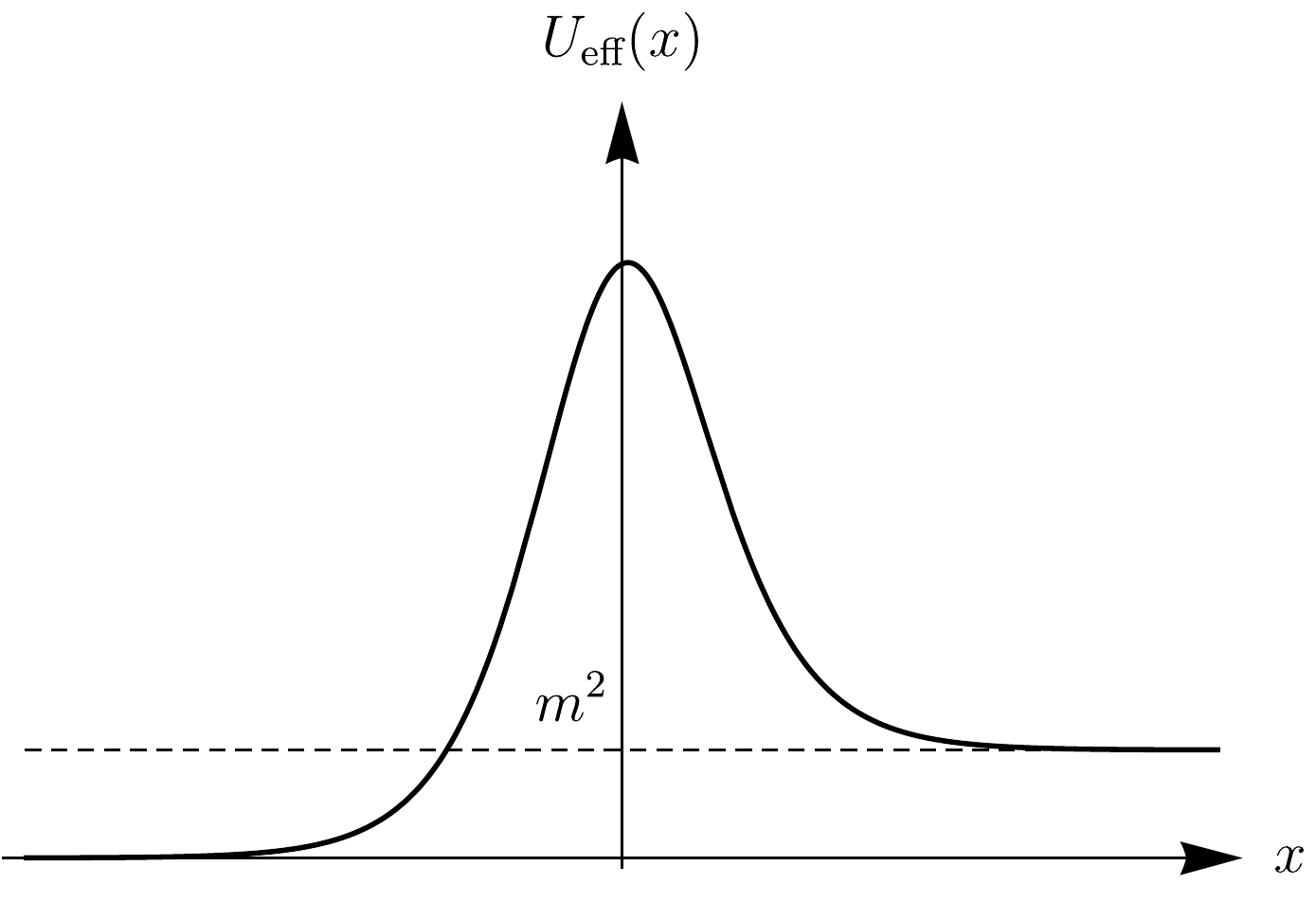}
    \caption{Potential for massive linear scalar modes in the
      two-dimensional dilaton BH background in the presence of the
      scalar-dilaton coupling with $q>m^2/(2\l^2)$. The horizon is
      located at $x\to-\infty$.} 
    \label{fig:Ueff}
\end{figure}

Let us study linear perturbations around the false vacuum $\vf=0$. To this end, we neglect the self-interaction part of the potential $V_{\rm int}$, and decompose $\vf$ using a complete set of positive- and negative-frequency modes:
\begin{equation}
\label{modes_gen}
    \vf_\o^+(t,x)=f_\o(x)\e^{-i\o t} \;, ~~~ \vf_\o^-(t,x)=f_\o^*(x)\e^{i\o t} \;, ~~~ \o>0 \;.
\end{equation}
The equation for $f_\o$ follows from the linearized field equation for $\vf$ and reads
\begin{equation}
\label{ModeEq}
    -f_\o''+U_{\rm eff}(x)f_\o=\o^2f_\o \;,
\end{equation}
where prime denotes derivative with respect to $x$. This is the
Schr\"{o}dinger equation with the potential 
\begin{equation}
\label{Ueff_gen}
    U_{\rm eff}(x)=m^2\O+\frac{2Q}{M}\O' \;.
\end{equation}
It is instructive to compare it with the analogous potential for
spherically-symmetric linear perturbations of the massive scalar field
in the four-dimensional Schwarzschild background, which we review in
appendix~\ref{app:Schw}. The latter admits
similar form as in \cref{Ueff_gen} (with a different $\O$), with the
factor in front of $\O'$ being proportional to the BH temperature.
To reproduce this behavior, in what
follows we take the mass of the dilaton BH to be temperature-dependent, 
\begin{equation}
\label{M(T)}
    M(\l)=\frac{M_0^2}{\l} \;,
\end{equation}
where $M_0$ is a constant of mass dimension 1.\footnote{The constant
  $M_0$ is subject to certain conditions ensuring that the back-reaction of
  vacuum decay on the background geometry is negligible; see appendix
  \ref{app:dilaton}. Apart from this, it is arbitrary.} In other
words, we enforce the relation between the BH mass and temperature as
in General Relativity. Using \cref{O}, we obtain 
\begin{equation}
\label{Ueff}
    U_{\rm eff}(x)=\frac{m^2}{1+\e^{-2\l x}}+\frac{2q\l^2\e^{-2\l x}}{(1+\e^{-2\l x})^2} \;,
\end{equation}
where we introduced $q=2Q/M_0^2$. The first term in this expression
describes a smooth interpolation between the near-horizon and
asymptotically-flat regions, while the second term generates a
barrier separating these regions, see Fig.~\ref{fig:Ueff}.  
 
Two comments are in order regarding the form of the potential
(\ref{Ueff}). 
First, at $2q\l^2>m^2$, the height of the barrier
exceeds $m^2$, and the maximum of the potential is achieved near
$x=0$, see Fig.~\ref{fig:Ueff}. This is expected to significantly
affect the properties of the tunneling solution in the vicinity of the BH.
Second, the mode
equation (\ref{ModeEq}) with the potential (\ref{Ueff}) admits a
general solution in terms of the hypergeometric function. This means,
in particular, that the Green's functions of $\varphi$ 
can be found analytically both near and far from the horizon. 
Thus, the linear part of the theory
(\ref{action_gen}) retains all good features of the model without the
scalar-dilaton coupling that was studied in
\cite{Shkerin:2021zbf}. Further properties of the 
potential (\ref{Ueff}) are discussed in
appendix \ref{app:Green}.

To match the greybody factor of the four-dimensional Schwarzschild
geometry, the parameter $q$ in \cref{Ueff} must be of the order of
one; see appendix \ref{app:Schw}. However, our goal is not to match
exactly the two-dimensional model with a spherical reduction of some
four-dimensional theory. Such a matching would anyway be imperfect due
to the difference in $\Omega$.
Instead, we aim at identifying
essential features of vacuum decay in four dimensions and modeling
them in the two-dimensional setup. 
As we believe that the presence of barrier is important, we expect
that the qualitative agreement between the physics of tunneling in the
two and four dimensions is achieved once $q$ exceeds $m^2/(2\l^2)$,
which at high temperature can be much less than one. 
We will see in secs.~\ref{sec:HH} and
\ref{sec:U} that it suffices to study the case 
\begin{equation}
\label{bound_p}
    m^2/(2\l^2)\lesssim q\ll 1\;.
\end{equation}
This limit significantly simplifies the calculation of
the Green's functions which is performed in appendix \ref{app:Green}.

\subsection{Bounce solution and tunneling rate}
\label{ssec:bounce}

\begin{figure}[t]
    \centering
    \includegraphics[width=0.45\linewidth]{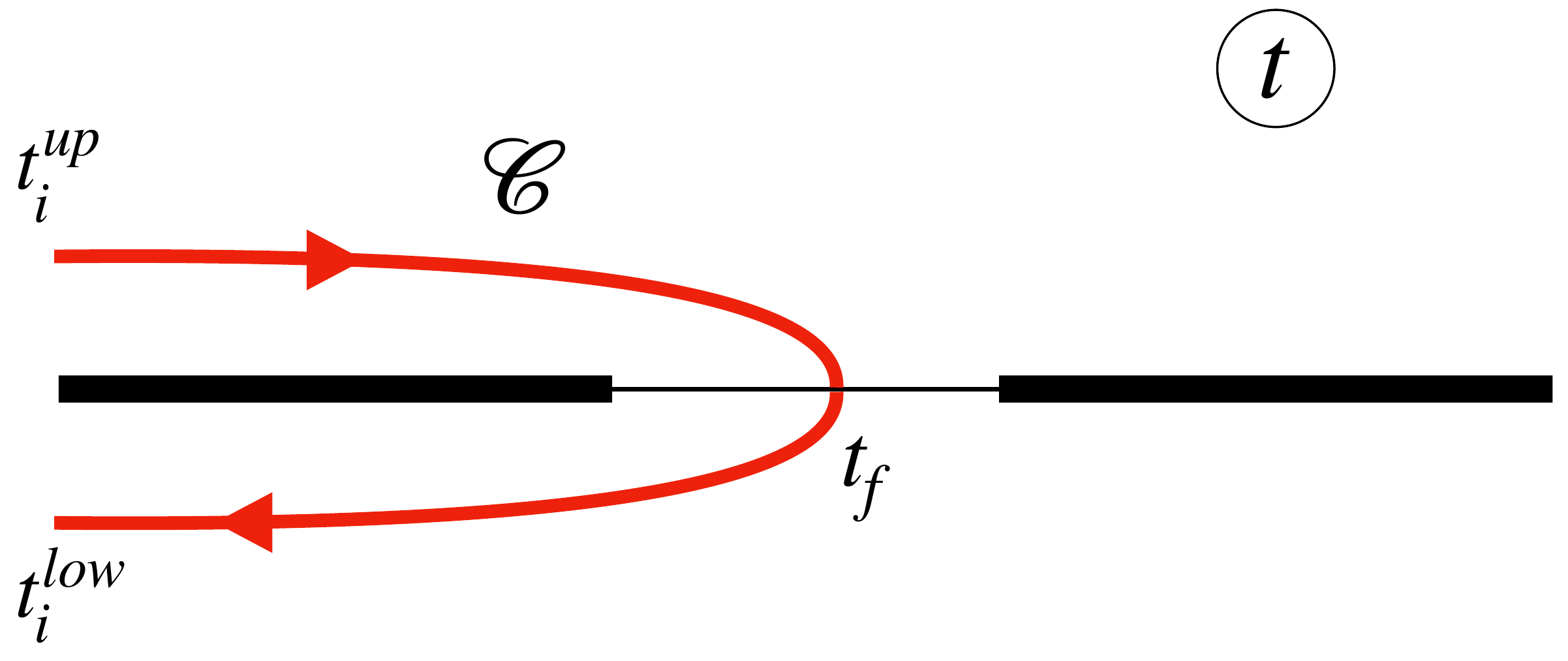}
    \caption{Contour $\C$ in the complex time plane for the
      calculation of the vacuum decay probability. We show the case
      when the branch-cuts of the bounce (shown with thick black
      lines) lie on the real axis. This corresponds to a theory with
      the scalar potential unbounded from below.}
    \label{fig:contour}
\end{figure}

Here we summarize the results of Ref.~\cite{Shkerin:2021zbf}
concerning the construction of bounce solutions. Bounce $\bb$ is a
regular solution of the field equations that saturates the transition
amplitude from the false vacuum initial state to the basin of
attraction of true vacuum. In our case, the field equation of motion
is 
\begin{equation}
\label{eom_b}
    \Box\vf-U_{\rm eff}(x)\vf-\O V_{\rm int}'(\vf)=0 \;.
\end{equation}
The bounce lives on a contour $\C$ in the complex time plane shown in
Fig.~\ref{fig:contour}. The
contour runs from the initial moment of time in the asymptotic past,
$t=t_i^{up}$ shifted to the upper half plane, to the final moment
$t=t_f$ and back to the asymptotic past, $t=t_i^{low}$ in the lower
half plane. The contour
must bypass the singularities of the bounce. Assuming that the bounce is
unique, its values on the upper and lower sides of the contour are
complex conjugated, hence at $t=t_f$ it is real and can be
analytically continued along the real time axis where it describes the
evolution of the field after tunneling.  

In the limits $t\to t_i^{up}$ and $t\to t_i^{low}$ the bounce must
satisfy boundary conditions imposed by the false vacuum state. These
boundary conditions turn out to be the same as for the time-ordered
Green's function in the corresponding vacuum $X$
\cite{Shkerin:2021zbf}. The latter is defined as a time-ordered
average of the linear field operators $\hat{\vf}$ in the state $X$, 
\begin{equation}
    \G_{X}(t,x;t',x')={}_{X}\langle
    T(\hat{\vf}(t,x)\hat{\vf}(t',x'))\rangle_X \;. 
\end{equation}
In turn, the field operator is constructed out of the complete set of
modes (\ref{modes_gen}) in the standard way
\cite{Birrell:1982ix}. Explicit expressions for the time-ordered
Green's functions in the Hartle--Hawking and Unruh vacua are presented
in appendix \ref{app:Green}.  

A Green's function satisfies the equation
\begin{equation}
\label{eom_G}
    \left(\Box-U_{\rm eff}(x)\right)\G_X(t,x;t',x')=i\delta(t-t')\delta(x-x') \;.
\end{equation}
Using this property, the field equation (\ref{eom_b}) can be recast
into the integral form. To select a particular solution---bounce---of
the integral equation, one should, first, adopt the time-integration
contour $\C$ in the complex time plane and, second, choose the 
Green's function corresponding to a particular false vacuum $X$. 
Thus, we arrive at 
\begin{equation}
\label{int_eq}
    \bb(t,x)=-i\int_{\C}\diff t'\int_{-\infty}^{\infty}\diff x'\,\G_{X}(t,x;t',x')\O(x')V_{\rm int}'(\bb(t',x')) \;.
\end{equation}
This form of the bounce equation will be useful in what follows.

Finally, let us discuss the tunneling rate $\Gamma$. The latter is
defined as the probability of tunneling per unit time. We are
interested in the main exponential dependence and write 
\begin{equation}
\label{DecayRate}
    \Gamma\sim \e^{-B} \;.
\end{equation}
The coefficient $B$ is the sum of the imaginary part of the bounce action
computed along the contour $\C$ and a boundary term representing the
initial-state contribution. One can show that the boundary term
cancels upon integration by parts in the action and one is left with
\cite{Shkerin:2021zbf} 
\begin{equation}
\label{B_gen}
    B=-\frac{i}{\gc^2}\int_{\C}\diff t\int_{-\infty}^{\infty}\diff x \left( \frac{1}{2}\bb V_{\rm int}'(\bb)-V_{\rm int}(\bb) \right) \;.
\end{equation}

\subsection{Tunneling in the inverted Liouville potential}
\label{ssec:Liouville}

In general, solving \cref{int_eq} (or \cref{eom_b} on the contour $\C$
with the vacuum boundary conditions) requires a numerical procedure. A
big simplification of the problem happens in theories where the
nonlinear core of the bounce, where it probes the true vacuum region,
is much smaller in size than the Compton wavelength of the free field
$\propto m^{-1}$. Then the source in the integral (\ref{int_eq}) is
essentially point-like, and the solution outside the core is simply
proportional to the Green's function $\G_X$. On the other hand,
the core itself can be found by neglecting the mass term in the field
equation (\ref{eom_b}). The full solution is constructed by matching
the long-distance asymptotics of the core with the short-distance
asymptotics of the Green's function. 

\begin{figure}[t]
    \centering
    \includegraphics[width=0.5\linewidth]{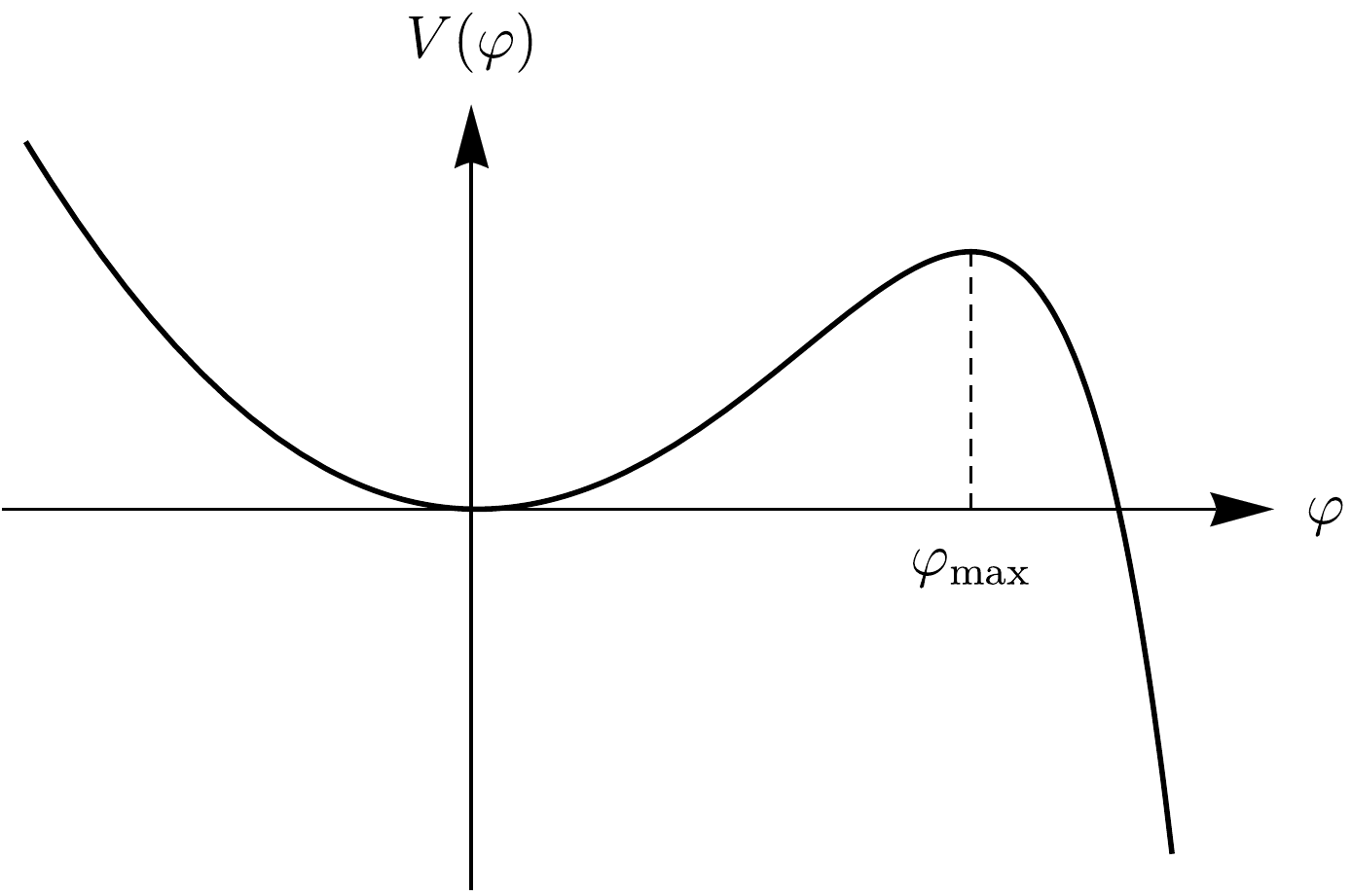}
    \caption{The scalar field potential.}
    \label{fig:pot}
\end{figure}

In \cite{Shkerin:2021zbf}, the interaction potential $V_{\rm int}$ has been studied for which the above procedure of finding the bounce solution works and yields analytic result. This is the negative Liouville potential,
\begin{equation}
\label{V_int}
    V_{\rm int}(\vf)=-2\k \left( \e^\vf-1 \right)
\end{equation}
with $\k>0$. In flat spacetime, the full scalar field potential $V(\vf)=\frac{1}{2}m^2\vf^2+V_{\rm int}(\vf)$ is shown in Fig.~\ref{fig:pot}. To ensure the applicability of the split-and-match procedure, the following relation between the parameters is adopted
\begin{equation}
\label{Hierarchy}
    \ln\frac{m}{\sqrt{\k}} \gg 1 \;.
\end{equation}
Thanks to this hierarchy, the theory possesses two intrinsic energy
scales: the mass scale $m$ and the scale associated with the maximum
of the scalar potential separating the false and true vacua
$m\ln\frac{m}{\sqrt{\k}}$. Generally speaking, the first
controls the width of the linear tail of the bounce, while the second
determines the size of its core. The maximum of the 
potential is located at 
\begin{equation}
\label{vf_max}
    \vf_{\rm max}\approx 2\ln\frac{m}{\sqrt{\k}} \;,
\end{equation}
where we have reatined only the leading logarithmic term.

\begin{figure}[t]
    \centering
    \includegraphics[width=0.45\linewidth]{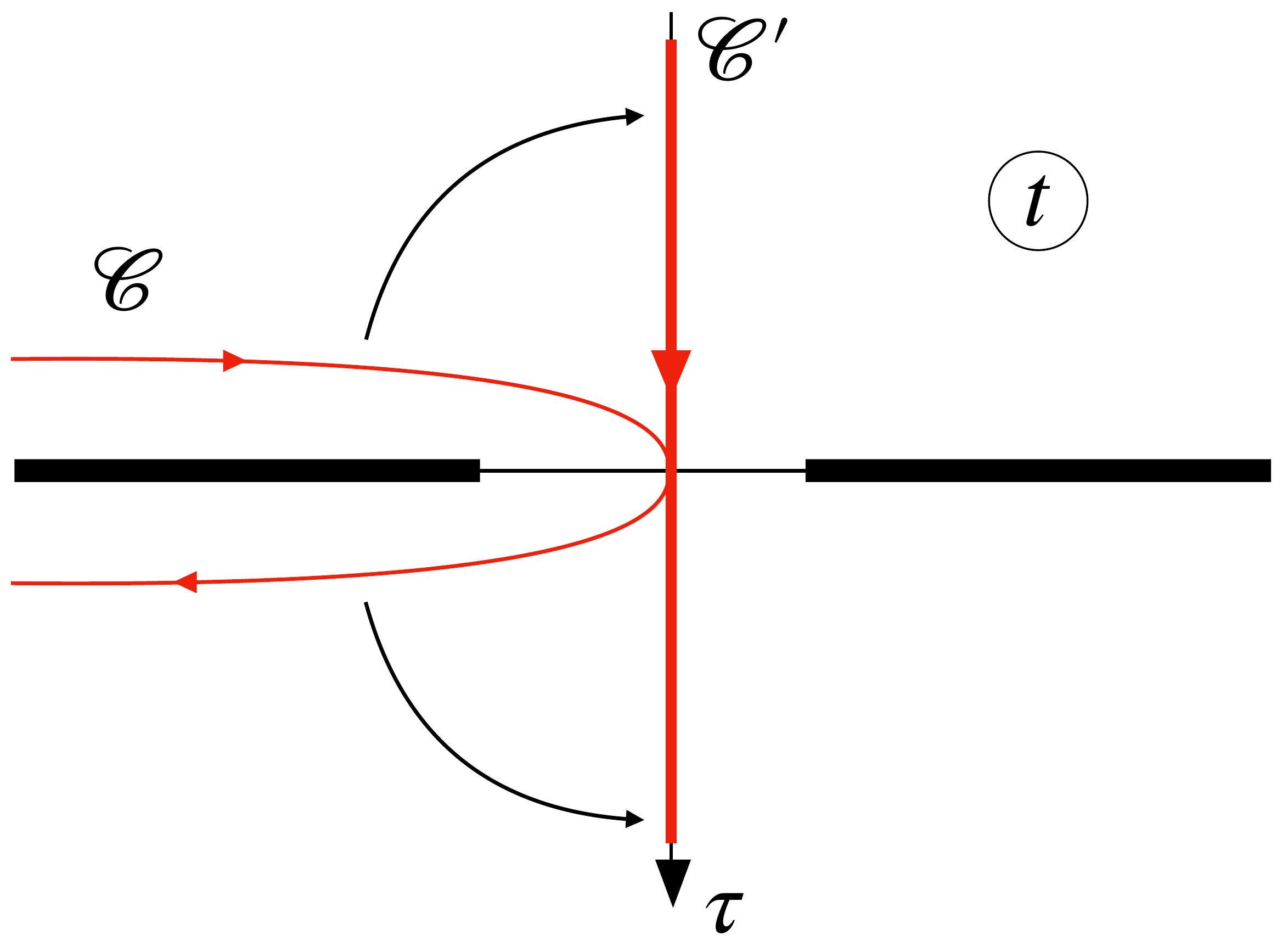}
    \caption{Deformation of the contour $\C$ into the Euclidean time
      contour $\C'$ used for the calculation of the flat-space
      Minkowski bounce. Black lines denote singularities of the
      bounce.} 
    \label{fig:contour_flat}
\end{figure}

To illustrate the matching procedure outlined above, let us discuss
bounce in the flat-space Minkowski vacuum (see
Ref.~\cite{Shkerin:2021zbf}). Assume that the only singularities of
the bounce are located on the real-time axis. Then, the contour $\C$
can be deformed into the contour $\C'$ that runs along the Euclidean
time axis, see Fig.~\ref{fig:contour_flat}. The vacuum boundary
condition at $\C$, which is provided by the Feynman Green's function,
becomes the vanishing boundary condition at $\C'$. Hence, the standard
Euclidean approach is reproduced \cite{Coleman:1977py,Callan:1977pt}.  

To find the core of the bounce, we neglect the mass term in
\cref{eom_b} and, using \cref{V_int}, obtain 
\begin{equation}
\label{eom_core_flat}
    \Box\left.\bb\right\vert_{\rm core} +2\k\e^{\left.\bb\right\vert_{\rm core}} =0 \;.
\end{equation}
This is the Liouville equation, and its general solution is
known. Next, the linear tail of the bounce is proportional to the
Feynman Green's function 
\begin{equation}
\label{Green_Feynman}
    \G_{F}(-i\tau,x;0,0)=\frac{1}{2\pi}K_0\left( m\sqrt{\tau^2+x^2+i\epsilon} \right) \;,
\end{equation}
where $\tau=it$ is the Euclidean time coordinate. The core and the
tail are matched in the region where, on the one hand, the solution to
the Liouville equation is linearized and, on the other hand, the
Green's function is approximated by its short-distance asymptotics. We
obtain\footnote{Solving the partial differential equation for the core
  (\ref{eom_core_flat}) becomes trivial once one takes into account
  that the solution providing the dominant decay channel is
  spherically-symmetric
  \cite{Coleman:1977th,Blum:2016ipp,0806.0299}. Note, however, that
  \cref{b_flat_core} can be derived without adopting spherical
  symmetry from the onset; in fact, the latter follows from the form
  of the Feynman Green's function (\ref{Green_Feynman}). } 
\bseq
\label{b_flat}
\begin{align} 
\label{b_flat_core}
&    \left.\bb\right\vert_{\rm core}=\ln\left[ \frac{4C_M^2}{(1+\k C_M^2(\tau^2+x^2))^2} \right] \\[1em]
\label{b_flat_tail}
&   \left.\bb\right\vert_{\rm tail}=8\pi\G_{F}(-i\tau,x;0,0) \;,
\end{align}
\eseq
where $C_M=m^2\e^{2\g_E}/(2\k)$. The matching region is defined by
\[
(C_M\sqrt{\k})^{-1}\ll\sqrt{\tau^2+x^2}\ll m^{-1}\;.
\] 
Its existence is
ensured by \cref{Hierarchy}. We see that the bounce is real for real
$\tau$, vanishes at infinity and has zero time derivative at
$\tau=0$. Hence, it is a valid tunneling solution. Moreover, it has no
singularities apart from the ones on the real-time axis, which
justifies the deformation of the contour $\C$ into $\C'$.  

The tunneling suppression is given by \cref{B_gen} where one should substitute the core of the bounce (\ref{b_flat_core}). We obtain\footnote{The corrections to \cref{B_M} are of order $\gc^{-2}\times o(1)$.}
\begin{equation}
\label{B_M}
    B_M=\frac{16\pi}{\gc^2}\left( \ln\frac{m}{\sqrt{\k}}+\g_E-1 \right) \;,
\end{equation}
where $\g_E$ is the Euler constant. We observe that the suppression is enhanced by the large logarithm (\ref{Hierarchy}).

In the BH background (\ref{O}), the linearized field equation
(\ref{ModeEq}) is still exactly solvable, and this allows us to
compute explicitly the Green's functions of interest. The solvability
of the equation for the core of the bounce is lost, but can be
recovered in the two regions: near the horizon where the metric is
approximately Rindler, and far away from the BH where the
spacetime is asymptotically flat. From \cref{O,Ueff} we see that the
two regions are defined by $x<0$, $|x|\gg \l^{-1}$ and $x>0$, $|x|\gg
\l^{-1}$, respectively. Although we do not have the explicit bounce
solution in the transition region $|x|\lesssim\l^{-1}$, we will still
be able to draw a qualitative picture of the evolution of the bounce
across this region.

\section{Decay of the Hartle--Hawking vacuum}
\label{sec:HH}

Here we study transitions from the Hartle--Hawking state in the model
defined by \cref{action_O,O,V_int}. This state corresponds
to the BH in thermal equilibrium, for which our method reduces to the standard Euclidean time approach.
It provides a benchmark for later study of transitions from the Unruh
vacuum. Besides, the relevant configuration describing vacuum decay
from the state in thermal equilibrium is readily found in the
four-dimensional BH background.
Comparing the results in two and four dimensions, we will
identify the regime in which our model adequately mimics catalysis
of false vacuum decay by the four-dimensional Schwarzschild BH. In
this and the following sections we will assume $\l\gg m$, which allows
us to treat the problem analytically.

\subsection{Tunneling in the black hole vicinity}
\label{ssec:HH_near}

\begin{figure}[t]
    \centering
    \includegraphics[width=0.45\linewidth]{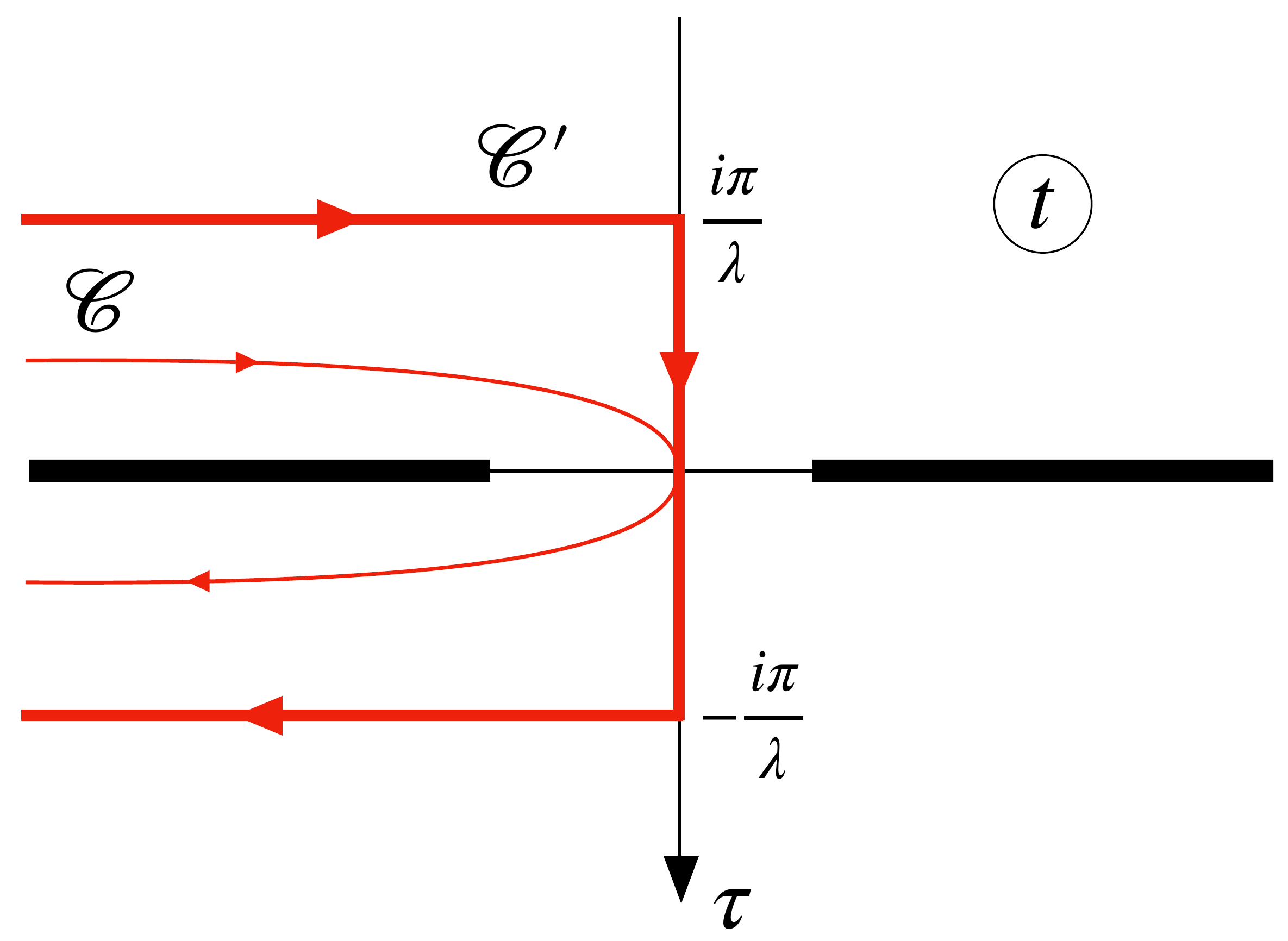}
    \caption{Deformation of the contour $\C$ into $\C'$ containing the Euclidean time segment which is used for the calculation of the Hartle--Hawking bounce.}
    \label{fig:contour_HH}
\end{figure}

Consider first the near-horizon region where the BH geometry is
approximated by the Rindler spacetime. The bounce solution is found by
applying the split-and-match procedure described in
sec.~\ref{ssec:Liouville}. It lives on the contour $\C$ stretched
along the real-time axis. The contour can be deformed partially to the
Euclidean time domain as shown in Fig.~\ref{fig:contour_HH}.  Then the
thermal boundary condition in the asymptotic past imposed at $\C$
transforms into the periodic boundary condition imposed at the
Euclidean segment $-\pi/\l<\tau<\pi/\l$ of $\C'$ (where $\tau=it$
denotes the Euclidean time coordinate) \cite{Shkerin:2021zbf}. 
This way one recovers the
standard Euclidean prescription for the thermal bounce
\cite{Linde:1981zj,Linde:1980tt,Ai:2018rnh}.  

The equation for the nonlinear core of the Hartle--Hawking bounce takes
the form (cf. \cref{eom_core_flat}) 
\begin{equation}
\label{eom_core_R}
    \Box\left.\bb\right\vert_{\rm core} +2\k\e^{2\l x + \left.\bb\right\vert_{\rm core}} =0 \;,
\end{equation}
where we assumed that the core fits the near-horizon region. This equation
admits analytic general solution. On the other hand, the linear tail of
the bounce is proportional to the time-ordered Hartle--Hawking Green's
function $\G_{HH}$ computed in the BH vicinity. Overall, we obtain
\cite{Shkerin:2021zbf} 
\bseq
\label{b_HH}
\begin{align}
\label{b_HH_core}
    & \left.\vf_{\rm b}\right\vert_{\rm core}=\ln\left[ \frac{\l^2b_{HH}}{\k\left(\ch\l(x-x_{HH})-\sqrt{1-b_{HH}}\cos\l\tau \right)^2} \right]-2\l x \\[1em]
\label{b_HH_tail}
    & \left.\vf_{\rm b}\right\vert_{\rm tail}=8\pi\G_{HH}(-i\tau,x;0,x_{HH})
\end{align}
\eseq
Here the parameter $b_{HH}$ is determined from matching the core with the tail, and $x_{HH}<0$ is the position of the center of the bounce.

Let us see how the core and the tail of the bounce match each
other. This is readily done if $b_{HH}\ll 1$, since in this case the
matching region exists in the Euclidean strip of the contour $\C'$. In
the matching region, on the one hand, the core of the bounce
linearizes, that is, the first term in the denominator of
$\left.\vf_{\rm b}\right\vert_{\rm core}$ dominates over the second
one, and, on the other hand, the tail of the bounce is taken in the
limit $|x-x_{HH}|\ll m^{-1}$. The expression for $\G_{HH}$ in this
limit is given in \cref{Green:GHH}. One obtains the following relation
between the parameters 
\begin{equation}
    \label{b_HH_b}
    b_{HH}=\frac{\k}{4\l^2}\e^{\frac{4\l}{m+q\l}-2\l x_{HH}} \;.
\end{equation}
By extending the matching region to the parts of the contour going
parallel to the real-time axis, one can show that \cref{b_HH_b}
remains valid as long as $b_{HH}\leq 1$ \cite{Shkerin:2021zbf}. On the other hand, no
matching is possible if $b_{HH}>1$. 

We see that the Hartle--Hawking bounce is characterized by one free
parameter $x_{HH}$, the position of the center of the
bounce. Existence of one-parameter family of solutions is a
consequence of the (approximate) Rindler symmetry of the BH
vicinity. The flat direction corresponding to $x_{HH}$ is tilted by
the terms in the BH metric that discriminate it from the Rindler
metric. 
Hence, one expects to get a unique tunneling solution with the
least suppression. The most likely candidate for such a solution is a
static sphaleron \cite{BH-3} (see also
\cite{Arnold:1989cq,Gregory:2013hja}). It is obtained in the limit
$b_{HH}=1$ which is achieved when $x_{HH}$ takes the value 
\begin{equation}
    x_{HH,\rm{sph}}=\frac{2}{m+q\l}-\frac{1}{\l}\ln\frac{2\l}{\sqrt{\k}} \;.
\end{equation}
The core of the sphaleron is given by
\begin{equation}
    \left.\vf_{\rm sph}\right\vert_{\rm core}=\ln\left[ \frac{\l^2}{\k\ch^2\left(\l(x-x_{HH,\rm{sph}}) \right)} \right] -2\l x \;.
\end{equation}
Note that, strictly speaking, the static sphaleron does not
satisfy the vacuum boundary conditions imposed at $\C$. 
Nevertheless,
as explained in \cite{Shkerin:2021zbf}, it appears as the end-point
configuration of valid tunneling solutions and correctly describes the
tunneling rate. Having this in mind, in the rest of this section we will focus on the sphaleron solution.

At low and moderate temperature, the coordinate of the sphaleron's center
is negative, $x_{HH,{\rm sph}}<0$, so it is comfortably inside the
near-horizon region. In the case without the dilaton barrier, 
$q=0$, the center of the sphaleron
shifts to the right as the temperature increases and at
$\l_{HH,0}\simeq \tfrac{m}{2}\ln\tfrac{m}{\sqrt\kappa}$ it reaches
$x=0$. At this point the sphaleron does not fit into the near-horizon
region anymore and sticks out into the flat space at $x>0$
\cite{Shkerin:2021zbf}. Our goal now is to understand what happens with the sphaleron at $\lambda\sim\lambda_{HH,0}$ in the presence of the barrier.

\subsection{Sphaleron at high temperature: weak and strong barriers}
\label{ssec:HH_far}

For static configurations, the equation of motion (\ref{eom_b}) reduces to the ordinary differential equation
\be
\label{spheq}
\vf''_{\rm sph}-(m^2\O+q\l\O')\vf_{\rm sph}+2\kappa \O\,\e^{\vf_{\rm
    sph}}=0\;, 
\ee
which is straightforward to solve numerically. It is instructive,
however, to consider a simplified version of \cref{spheq} which
can be studied analytically. To this purpose, we replace the function
$\O(x)$ in the brackets by the Heaviside $\theta$-function and its
derivative by the $\delta$-function. Physically, this means that we
are neglecting the width of the dilaton barrier compared to
the size of the 
sphaleron. This is certainly a good approximation for the sphaleron
tails which have width of order $m^{-1}$. On the other hand, the
sphaleron core has width $\sim \l^{-1}$ which is comparable to the
width of the barrier. Below we will encounter configurations with the
core of the sphaleron in the immediate neighborhood of the
barrier. For these configurations we do not expect an exact
quantitative agreement with the solution of the original
\cref{spheq}. Nevertheless, we will see that they capture the
right qualitative behavior. 

We also simplify the last --- Liouville --- term in \cref{spheq}. We
cannot simply set $\O$ to the $\theta$-function in it as this would
lead to the loss of sphaleron solutions in the near-horizon
region. Instead, we approximate $\O$ with a pure exponential $\e^{2\l
  x}$ at $x<0$ and $1$ at $x>0$. In other words, we assume that the
metric is exactly Rindler to the left from the barrier and flat to the
right of it. Overall, the approximate equation we will analyze has the
following form, 
\be
\label{spheq1}
\vf''_{\rm sph}-\big(m^2\theta(x)+q\l\delta(x)\big)\vf_{\rm sph}
+2\kappa \big(\theta(-x)\e^{2\l x}+\theta(x)\big)\e^{\vf_{\rm
    sph}}=0\;. 
\ee  
It is straightforward to solve this equation to the left and to the
right of the barrier. The solutions bounded at $x\to\pm\infty$ are:
\begin{align}
\label{sphL}
&\vf_{\rm sph}\Big|_{\rm
  left}=\ln\bigg[\frac{\l^2}{\kappa\ch^2(\l(x-x_{\rm sph,L}))}
\bigg]-2\l x\;,\\
~\notag\\
\label{sphR}
&\vf_{\rm sph}\Big|_{\rm right}=
\begin{cases}
\ln\bigg[\frac{\l_0^2}{\kappa\ch^2(\l_0(x-x_{\rm sph,R}))}
\bigg]~, & \text{core}\\
\frac{2\l_0}{m}\e^{-m|x-x_{\rm sph,R}|}~,& \text{tail}
\end{cases}
\end{align}
where $\l_0$ satisfies
\be
\label{l0eq}
\frac{\l_0}{\ln(2\l_0/\sqrt\kappa)}=m ~~~ \Rightarrow ~~~ \l_0\simeq
m\ln\frac{m}{\sqrt{\k}}\;. 
\ee
At $x=0$ the two solutions
(\ref{sphL}), (\ref{sphR}) must be matched continuously, whereas the
derivative must have a jump due to the $\delta$-function in the
equation,
\bseq
\label{HHbcs}
\begin{align}
\label{HHbc1}
&\vf_{\rm sph}(0)\Big|_{\rm right}=\vf_{\rm sph}(0)\Big|_{\rm
  left}\equiv \vf_{\rm sph}(0)\;,\\
\label{HHbc2}
&\vf'_{\rm sph}(0)\Big|_{\rm right}=\vf'_{\rm sph}(0)\Big|_{\rm
  left}+q\l \vf_{\rm sph}(0)\;.
\end{align}
\eseq

Let us assume first that the solution on the right is given purely by
the tail, i.e., the core of the sphaleron lies deep in the near-horizon
region. Then the matching conditions (\ref{HHbcs}) lead to the
equation on $x_{\rm sph,L}$,
\be
\label{xsphL}
1-\th(\l x_{\rm sph,L})=\frac{m+q\l}{2\l} \ln\bigg[\frac{\l^2}{\kappa
  \ch^2(\l x_{\rm sph,L})}\bigg]\;.
\ee 
To proceed, it is convenient to parameterize $q$ as
\be
\label{qa}
q=\frac{a}{\ln(m/\sqrt\kappa)}\;.
\ee
For $a>1$, \cref{xsphL} has a negative solution for arbitrary
value of $\l\gg m$, implying that the sphaleron core is always confined to
the near-horizon region. This behavior is dramatically different from
the case without barrier ($a=0$) and we will call barriers with $a>1$
``strong''. Notice that even for a strong barrier $q$ itself can be
much smaller than $1$. 

If $a<1$, the negative solution to \cref{xsphL} grows with
temperature and reaches zero at $\l\simeq
\tfrac{m}{1-a}\ln\tfrac{m}{\sqrt\kappa}$. The sphaleron stops fitting
the near-horizon region and escapes outside. We will refer to this
case as ``weak barrier''. When $x_{\rm sph,L}$ gets close to zero,
\cref{xsphL}  becomes inaccurate because the solution on the right
can no longer be approximated by a pure tail. Instead, the sphaleron
core is now sitting right on the barrier and we should use the upper
expression in \cref{sphR}. Let us introduce
\be
\label{sLsR}
s_{\rm L}=\l \th(\l x_{\rm sph,L})~,~~~~~
s_{\rm R}=\l_0 \th(\l_0 x_{\rm sph,R})\;.
\ee
Note that we do not require $x_{\rm sph,R}$ ($x_{\rm
  sph,L}$) to be positive (negative) --- these are just the parameters
of the solution and can have either sign. The matching conditions (\ref{HHbcs})
take the form
\bseq
\label{sseq}
\begin{align}
\label{sseq1}
&\l^2-s_{\rm L}^2=\l_0^2-s_{\rm R}^2\;,\\
\label{sseq2}
&s_{\rm L}-s_{\rm R}=\l(1-a)\;,
\end{align}
\eseq
where in the second line we substituted \cref{qa} and neglected
terms suppressed by the large logarithm (\ref{Hierarchy}). The
solution reads 
\be
\label{sssol}
s_{\rm L}=\frac{\l^2(2-2a+a^2)-\l_0^2}{2\l(1-a)}~,~~~~~
s_{\rm R}=\frac{\l^2 a(2-a)-\l_0^2}{2\l(1-a)}\;.
\ee
The conditions $-1<s_{\rm L}/\l,s_{\rm R}/\l_0<1$ following from the
definition (\ref{sLsR}) are satisfied for BH
temperatures in the interval
\be
\label{HHlint}
\frac{\l_0}{2-a}<\l<\frac{\l_0}{a}\;,
\ee
which is non-empty only for $a<1$. At the lower end of this interval
both $x_{\rm sph,L}$ and $x_{\rm sph,R}$ formally go to $-\infty$,
which means that the solution matches to the sphalerons deep in the
near-horizon region studied above (the run-away is regularized by
replacing the sphaleron core on the right by the tail).  

The behavior at the upper end is qualitatively different for the case
without barrier ($a=0$) and with barrier, no matter how weak. In the
former case $x_{\rm sph,R}\to 0$ at $\l\to \infty$ implying that the
field at $x>0$ represents half of the flat-space sphaleron. At the
same time $x_{\rm sph,L}\to +\infty$, which means that the field at
$x<0$ is simply constant. Thus, the two-dimensional BH ``cuts the
sphaleron in half'' \cite{Shkerin:2021zbf}. Accordingly, the sphaleron
energy is half of that in flat space. 

On the other hand, for $a>0$ both $x_{\rm sph,R}$ and $x_{\rm sph,L}$
run away to $+\infty$ at $\l\to \l_0/a$. This means that the sphaleron
core gets detached from the barrier and shifts into the flat region to
the right. At $\l>\l_0/a$ there are no sphalerons at finite distance
from the barrier. Of course, there are still flat-space sphalerons in
the asymptotic region $x\to +\infty$ given by \cref{sphR} and the
Hartle--Hawking vacuum decay proceeds via jumps over these sphalerons
in the thermal bath far away from the BH.

\begin{figure}[t]
    \centering
   \includegraphics[width=0.6\linewidth]{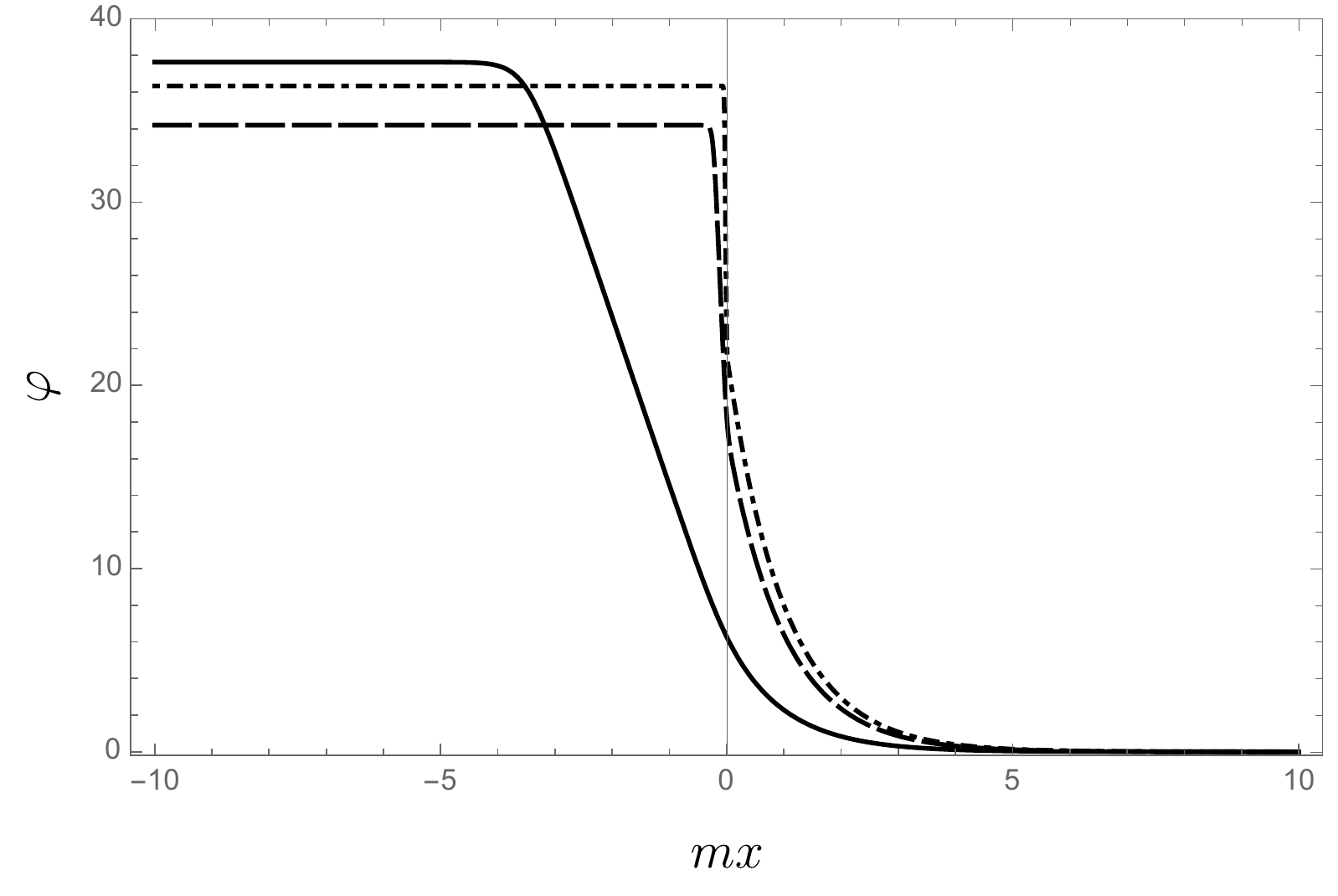}
    \caption{Profiles of the sphaleron in the metric of
      two-dimensional dilaton BH with strong dilaton
      barrier. We take $\k=10^{-8}m^2$ and
      $q=1.5/\ln\tfrac{m}{\sqrt{\k}}$. Different curves correspond to the
      following BH temperatures: $\l/\big(m\ln\tfrac{m}{\sqrt{\k}}\big)=0.25$
      (solid), $2$ (dashed), $10$
      (dot-dashed).}
    \label{fig:strong}
\end{figure}

\begin{figure}[t]
    \centering
    \includegraphics[width=0.6\linewidth]{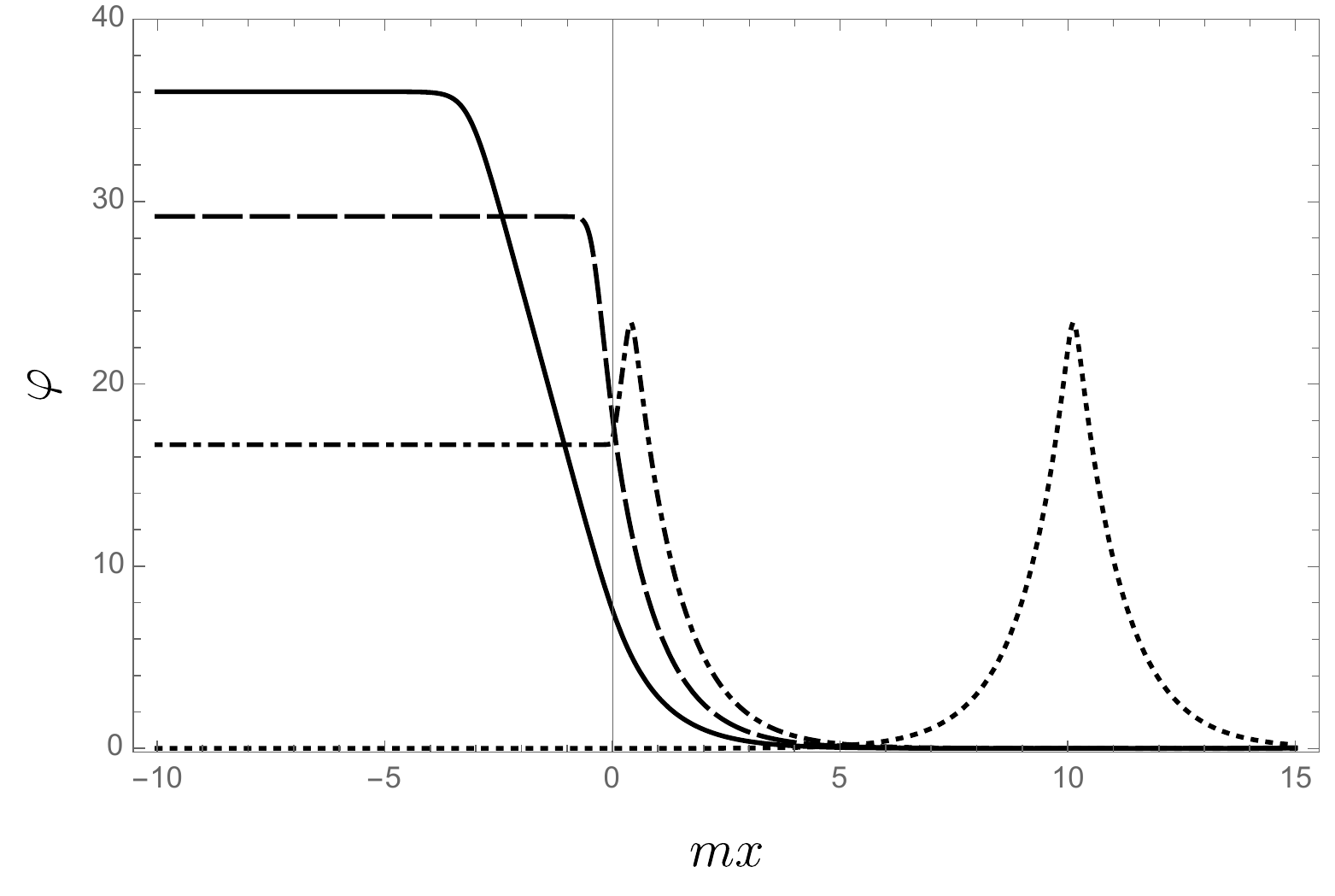}
    \caption{Same as Fig.~\ref{fig:strong}, but now for the case of 
      weak dilaton
      barrier. We take $\k=10^{-8}m^2$ and
      $q=0.5/\ln\tfrac{m}{\sqrt{\k}}$. The BH temperatures are: 
$\l/\big(m\ln\tfrac{m}{\sqrt{\k}}\big)=0.25$
      (solid), $0.66$ (dashed), $1.942$ (dot-dashed), $1.944$
      (dotted).}
    \label{fig:sph_field}
\end{figure}

The above analysis is confirmed by the direct numerical solution of
\cref{spheq}. In Figs.~\ref{fig:strong} and \ref{fig:sph_field}
we plot the sphaleron profiles for several values of BH temperature
for strong ($a=1.5$) and weak ($a=0.5$) barriers. We see that for the
strong barrier the sphaleron gets confined to the near-horizon region
at all temperatures. By contrast, in the weak barrier case, as the BH
temperature increases, the
sphaleron shifts from the near-horizon region to be centered on 
the barrier and then moves
further out to the asymptotically flat region.

Which of the two regimes---the weak barrier or the strong
barrier---corresponds better to a realistic four-dimensional theory?
To answer this question, in appendix \ref{app:four} we perform a
numerical investigation of the decay of the Hartle--Hawking vacuum in
the Schwarzschild background in four dimensions. We focus again on the
sphaleron solution since one can argue that it is the most relevant configuration at all BH temperatures \cite{BH-3}. We take the
theory of a massive scalar field with a negative quartic
self-interaction, which is a prototypical model of the Higgs field and
its (loop-corrected) potential. Our analysis shows that the
high-temperature Hartle--Hawking sphaleron tends asymptotically to its
flat-space counterpart. This means that there are no solutions
localized in the near-horizon region of a small Schwarzschild BH in thermal
equilibrium. We conclude that the case of weak barrier describes more
adequately the physics of vacuum decay in four dimensions. Since our goal
is to model the four-dimensional physics as closely as possible, we
focus on the weak barrier case $0<a<1$ in the rest of the paper.

Let us comment on the value of the scalar-dilaton coupling,
$q=(\ln\frac{m}{\sqrt{\k}})^{-1}\ll 1$, separating the regimes of weak
and strong barrier. That in our model this value is much less than 1
is due to the large hierarchy between the mass scale and the scale of
the scalar potential separating the false and true vacua, see
\cref{Hierarchy,vf_max}. This leads to the hierarchy between the mass
and the temperature at which the core of the sphaleron reaches
outside, $\l_{HH,0}\gg m$. To prevent the sphaleron from escaping
the BH vicinity, the barrier must be such that $q\l_{HH,0}\gtrsim m$
which is already achieved at $q\ll 1$. The situation is different in
the four-dimensional model studied in appendix \ref{app:four}, since
there the sphaleron core is of the size of the Compton wavelength and
reaches outside at $\l\sim m$. Hence, the value $q\sim 1$ corresponding
to the Schwarzschild background is still not enough to confine the
solution.

\subsection{Decay probability}

Let us now discuss the decay rate. We start from the low-temperature
case when the bounce forms in the near-horizon region. The expression
for the suppression of the Hartle--Hawking bounce for general values
of $b_{HH}\leq 1$ and $x_{HH}<0$ was derived in \cite{Shkerin:2021zbf}
and reads,\footnote{Corrections to this formula are of order ${\rm
    g}^{-2}\times o(1)$.}
\be
\label{BHHlow}
B_{HH,\text{low-}\l}=\frac{4\pi}{{\rm
    g}^2}\bigg(\ln\bigg[\frac{4\l^2}{\kappa b_{HH}}\bigg]-2\l
x_{HH}-4\bigg)\;. 
\ee  
The derivation is insensitive to the shape of the 
potential for linear modes 
in the transition region $|x|\lesssim \l^{-1}$ and, hence,
this formula
is readily applicable to our model. Using \cref{b_HH_b}, we obtain
\be
\label{BHHlow1}
B_{HH,\text{low-}\l}=\frac{16\pi}{{\rm
    g}^2}\bigg(\ln\frac{\l}{\sqrt\kappa}-\frac{\l}{m+q\l}\bigg)\;,
\ee
where, for simplicity, we have kept only the logarithmically enhanced
terms. This expression is valid as long as the core of the bounce or
sphaleron fits entirely into the near-horizon region, i.e., as long as 
\be
\label{lHH1}
\l\lesssim \l_{HH,1}\equiv \frac{m}{2-a}\ln\frac{m}{\sqrt\kappa}\;,
\ee
where $a$ is defined in \cref{qa}.

At higher temperature the vacuum decay proceeds via thermal jumps over the
sphaleron which is the saddle-point of the energy barrier separating
the 
vacua. The corresponding Boltzmann 
suppression is
\be
\label{Boltzmann}
B_{HH,\text{high-}\l}=\frac{2\pi E_{\rm sph}}{\l}\;.
\ee  
To find the sphaleron energy $E_{\rm sph}$, we rewrite the general expression for it using
integration by parts and equation of motion (\ref{spheq}):
\be
\label{Esph1}
\begin{split}
E_{\rm sph}&=\frac{1}{{\rm g}^2}\int_{-\infty}^{+\infty}\diff x
\bigg(\frac{1}{2}\vf'^2_{\rm sph}+\frac{1}{2}(m^2\O+q\l\O')\vf^2_{\rm
  sph} -2\kappa \O (\e^{\vf_{\rm sph}}-1)\bigg)\\
&=\frac{1}{{\rm g}^2}\int_{-\infty}^{+\infty}\diff x \,\kappa\O \Big((\vf_{\rm
sph}-2)\e^{\vf_{\rm sph}}+2\Big)\;.
\end{split}
\ee
Next, we substitute here the solution from the previous subsection
(\ref{sphL}), (\ref{sphR}) in the case when the sphaleron sits on the
dilaton
barrier and use \cref{qa,sssol}. Keeping only the leading logarithmically
enhanced terms, we obtain 
\be
\label{Esph2}
E_{\rm sph}\simeq\frac{2(m+q\l)}{{\rm g}^2}
\Big(\ln\frac{m}{\sqrt\kappa}\Big)^2\;.
\ee
This
gives us the sphaleron energy up to 
\be
\label{lHH2}
\l\lesssim \l_{HH,2}\equiv
\frac{m}{q}=\frac{m}{a}\ln\frac{m}{\sqrt\kappa}\;.  
\ee

As discussed before, at yet higher temperatures there are no
sphalerons at finite distance from the BH. The height of the energy
barrier 
separating the false and true vacua is then given by the energy of the
flat-space sphaleron in the asymptotic region $x\to +\infty$, which is
obtained by substituting (\ref{sphR}) into (\ref{Esph1}) and taking
the limit $x_{\rm sph,R}\to+\infty$. 
This yields,
\be
\label{Esphflat}
E_{\rm sph}^{\rm flat}\simeq \frac{4m}{{\rm g}^2}
\Big(\ln\frac{m}{\sqrt\k}\Big)^2\;.
\ee 

Substituting these results into \cref{Boltzmann}, we arrive at 
\be
\label{BHHhigh}
B_{HH,\text{high-}\l}\simeq\begin{cases}
\frac{4\pi(m+q\l)}{{\rm g}^2\l}
\left(\ln\frac{m}{\sqrt\kappa}\right)^2\;,& \l_{HH,1}<\l<\l_{HH,2}\\
\frac{8\pi m}{{\rm g}^2\l}
\left(\ln\frac{m}{\sqrt\kappa}\right)^2\;, & \l_{HH,2}<\l
\end{cases}
\ee
At $\l=\l_{HH,1}$ this expression matches smoothly (both the function
and its first derivative) to the low-temperature suppression
(\ref{BHHlow1}). The suppression at this temperature is one-half of
the suppression for tunneling in Minkowski spacetime (\ref{B_M}). On
the other hand, at $\l=\l_{HH,2}$ the suppression has a break
(although it is still continuous). This is due to the abrupt run-away
of the sphaleron to infinity. 

\begin{figure}[t]
    \centering
    \includegraphics[width=0.6\linewidth]{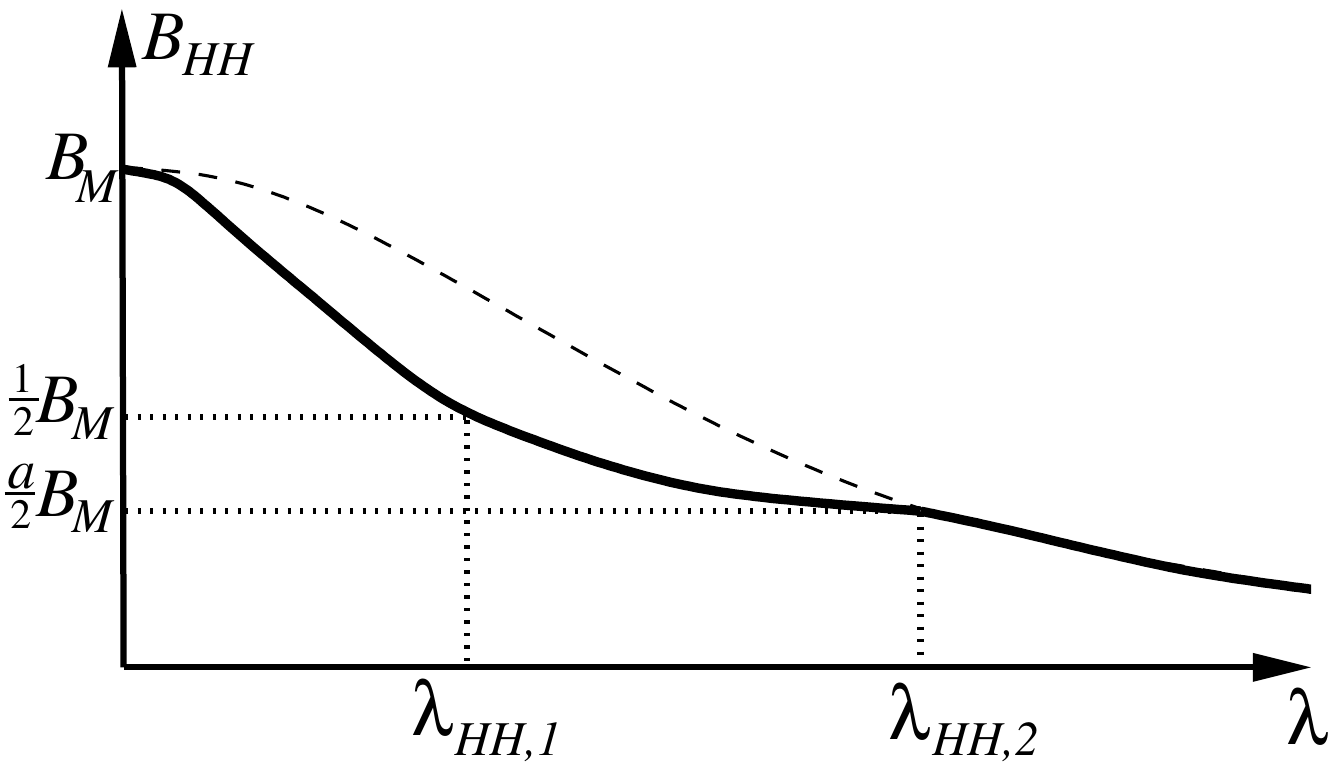}
    \caption{Exponential suppression of the Hartle--Hawking vacuum
      decay as a function of BH temperature $T_{BH}=\l/(2\pi)$ 
(solid) vs. suppression of
      the vacuum decay in a thermal bath with the same temperature in
      the absence of BH (dashed). At low temperature $\l<\l_{HH,1}$
      decay proceeds via tunneling in the near-horizon region, at
      $\l_{HH,1}<\l<\l_{HH,2}$ via sphaleron transitions in the BH
      vicinity, 
and at $\l_{HH,2}<\l$ via sphaleron transitions far
      from BH. The critical temperatures marking
      the boundaries of different regimes are given by
      \cref{lHH1,lHH2}. We work in the regime of weak dilaton barrier.}
    \label{fig:B_HH_weak}
\end{figure}

We plot the suppression of the Hartle--Hawking vacuum decay in
Fig.~\ref{fig:B_HH_weak}, where we compare it to the suppression of
decay in thermal bath at the same temperature in the absence of BH. The latter was calculated in \cite{Shkerin:2021zbf} and reads
\be
\label{Bthflat}
B_{\rm th}^{\rm flat}\simeq\begin{cases}
\frac{16\pi}{{\rm
    g}^2}\left(\ln\frac{\l}{\sqrt\kappa}-\frac{\l}{2m}\right)
\;,&\l<\l_0\\
\frac{8\pi m}{{\rm g}^2\l}\left(\ln\frac{m}{\sqrt\kappa}\right)^2\;,&
 \l_0<\l
\end{cases}
\ee  
where $\l_0$ is defined in \cref{l0eq}. We see that the BH provides an
additional enhancement of the decay rate in a range of temperatures,
but in the high temperature limit the catalyzing effect due to curved
geometry disappears. Of course, at $\l\to\infty$ the suppression
vanishes, as expected in a thermal state. 

Let us make an observation that will be important in what follows. The
sphaleron transition rate can be found using a simple stochastic
picture \cite{Shkerin:2021zbf}. At high temperature the occupation
numbers of the low-lying modes are large and the field experiences
large --- essentially classical --- fluctuations with a long correlation
length $\sim m^{-1}$. From time to time these fluctuations will throw the
field over the barrier separating the false and true vacua. The rate
of such events can be estimated as 
\be
\label{GammaHH}
\Gamma_{HH,\text{high-}\l}\sim\exp\bigg(-\frac{\vf_{\rm
    max}^2}{2(\delta\vf)^2_{HH}}\bigg)\;, 
\ee
where $\vf_{\rm max}$ is the value of the field at the maximum of the
scalar potential (\ref{vf_max}) and $(\delta\vf)_{HH}^2$ is the
variance of the fluctuations. In deriving \cref{GammaHH} we assumed
that the fluctuations are Gaussian. This is a good approximation in
our model since the scalar potential is essentially quadratic at
$\vf<\vf_{\rm max}$ and hence the field is free as long as its
amplitude does not exceed $\vf_{\rm max}$. 

The variance $(\delta\vf)_{HH}^2$ at the position $x$ 
can be found from the coincidence
limit of the Green's function, upon subtraction of the Green's
function in empty space to remove the ultraviolet divergence,
\be
\label{varHH}
(\delta\vf)_{HH}^2={\rm g}^2\lim_{t\to 0,\,x'\to
  x}\big[\G_{HH}(t,x';0,x)-\G_{F}(t,x';0,x)\big]\;.
\ee 
From the expressions for the Hartle--Hawking Green's function to the
right of the barrier, eq.~(\ref{Green:GHH_right}), 
we see that the variance changes between 
\be
\label{varHHnear}
(\delta\vf)_{HH}^2\simeq \frac{{\rm g}^2\l}{2\pi(m+q\l)}~~~~\text{at}~x=0
\ee
and 
\be
\label{varHHfar}
(\delta\vf)_{HH}^2\simeq \frac{{\rm g}^2\l}{2\pi m}~~~~\text{at}~x\to +\infty
\;.
\ee
Note that for $\l<\l_{HH,2}$ the fluctuations are larger in the
vicinity of the BH which implies that the transitions will
predominantly occur in this vicinity. However, for $\l>\l_{HH,2}$ the
amplitude of fluctuations near the BH is suppressed compared to
fluctuations at infinity. This is an effect of the high dilaton
barrier for modes at $x=0$ --- the field gets repelled from the BH. In
this case the transitions at infinity will be preferred, which confirms
our earlier findings about the absence of sphaleron at finite $x$ at
these temperatures.   

Substituting the maximal of the two expressions (\ref{varHHnear}),
(\ref{varHHfar}) into \cref{GammaHH}, we obtain
\be
\label{GammaHH1}
\Gamma_{HH,\text{high-}\l}\sim\begin{cases}
\exp\bigg[-\frac{4\pi(m+q\l)}{{\rm
    g}^2\l}\Big(\ln\frac{m}{\sqrt\kappa}\Big)^2\bigg]\;,&\l<\l_{HH,2}\\
 \exp\bigg[-\frac{8\pi m}{{\rm
    g}^2\l}\Big(\ln\frac{m}{\sqrt\kappa}\Big)^2\bigg]\;,&\l>\l_{HH,2}
\end{cases} 
\ee
which coincides with \cref{BHHhigh}. The stochastic approach is
particularly useful when construction of the actual semiclassical solution
is problematic. In the next section we will use it to estimate the
rate of Unruh vacuum decay at high temperature.

\section{Decay of the Unruh vacuum}
\label{sec:U}

We now turn to the main topic of this work --- decay of the Unruh
vacuum. We focus on the case of weak dilaton barrier, $a<1$, where $a$ is
defined in \cref{qa}. As explained in sec.~\ref{sec:HH}, in this
regime the transitions from the Hartle--Hawking state proceed in close
analogy with the four-dimensional case. It is reasonable to expect
that the same is true for the Unruh vacuum.   

Far from the BH, the Unruh vacuum corresponds to a flux of
thermal radiation whose spectrum is reduced by the temperature-dependent
barrier. Hence, we expect transitions far from BH 
to be more suppressed than in the model
without the barrier studied in \cite{Shkerin:2021zbf}. Close to the
horizon, by analogy with the Hartle--Hawking case, we expect that the
Unruh bounce exists in a certain range of BH temperatures and
disappears at a high enough temperature.

\subsection{Tunneling near horizon}
\label{ssec:U_near}

To find the Unruh bounce in the BH vicinity, we have to solve
\cref{eom_b} with $\O=\e^{2\l x}$ along the contour $\C$ shown in
Fig.~\ref{fig:contour} and satisfy the boundary condition imposed by
the Unruh vacuum.\footnote{As in the Hartle--Hawking case, $\C$ can be
deformed into the contour $\C'$ with a part in the Euclidean time domain, see
Fig.~\ref{fig:contour_HH}. This facilitates the
matching procedure and is legitimate since the deformation does not
intersect singularities of the bounce. Note, however, that unlike the
Hartle--Hawking case, the Unruh bounce is not real in Euclidean time.}  
In the core region of the bounce the field equation reduces to
\cref{eom_core_R}. The tail of the bounce is determined by the
time-ordered Unruh Green's function $\G_U$ computed in the BH
vicinity. Overall, we have \cite{Shkerin:2021zbf} 
\bseq
\label{b_U}
\begin{align}
\label{b_U_core}
    & \left.\vf_{\rm b}\right\vert_{\rm core}=\ln\left[
      \frac{4\l^2b_{U}}{\k\left(
          -2\l(v-x_U)\sh\left(\frac{\l}{2}(u+x_U)\right) +
          b_{U}\e^{\frac{\l}{2}(u+x_U)}\right)^2} \right]-2\l x
    \\[1em] 
\label{b_U_tail}
    & \left.\vf_{\rm b}\right\vert_{\rm tail}=8\pi\G_{U}(t,x;0,x_U)
\end{align}
\eseq
Here we defined $u=t-x$, $v=t+x$. 
The core expression (\ref{b_U_core}) has been built in such a way 
that in the linearized regime its $u$- and $v$-dependent
parts match the corresponding terms in the Green's function. The
parameter $x_U$ determines the 
position of the center of the bounce and $b_{U}$ is found from
matching the constant parts of (\ref{b_U_core}) and
(\ref{b_U_tail}). For simplicity, we will keep in this matching only
the terms enhanced by the large ratio $\l/m$. This yields
\begin{equation}
\label{bU1}
    b_{U}=\Bar{b}_{U}\e^{-2\l x_U} \;, ~~~~~ 
\Bar{b}_{U}=\frac{\k}{4m^2}\exp
\bigg\{\frac{4\l}{\pi m}\,\H\Big(\frac{q\l}{m}\Big)\bigg\}\;,
\end{equation}
where the function $\H$ is defined in \cref{Green:HL} and plotted in
Fig.~\ref{fig:Hs}.  
As in the Hartle--Hawking case, we obtain a one-parameter family of
solutions parameterized by $x_U$. Again, this is an artifact of our
approximation of the near-horizon geometry by the Rindler
spacetime. One expects that taking into account the deviation of BH
metric from Rindler will remove the degeneracy \cite{Shkerin:2021zbf}.  

For the applicability of the matching procedure the bounce core should
be smaller than the tail, which amounts to the requirement 
\begin{equation}
\label{bound_U1}
    \Bar{b}_{U}\lesssim 1 \;.
\end{equation}
More careful matching conditions can be found in
\cite{Shkerin:2021zbf}, but they are not important for what follows. 
Inequality (\ref{bound_U1}) translates into the upper bound on the BH
temperature, 
\be
\label{lambdaU}
\l<\l_{U}\equiv \frac{y_c m}{a}\ln\frac{m}{\sqrt\kappa}\;,
\ee
where $y_c$ is the solution of the equation
\be
\label{yc}
y_c\H(y_c)=\frac{\pi a}{2}\;.
\ee
Note that this solution exists for $a<1$ due to the asymptotics of the
function $\H$, eqs.~(\ref{Green:HL_as1}), (\ref{Green:HL_as2}). 
At
$\l>\l_{U}$, the bound (\ref{bound_U1}) is violated and the matching
procedure breaks down. At the same time, the Liouville core of the
Unruh bounce stops fitting the near-horizon region. By analogy with
the Hartle--Hawking case, we expect that further growth of
temperature drives the vacuum decay site across the gravitational
barrier and to the flat-space region on the right. We were not able to
find explicitly the corresponding bounce solutions. Nevertheless, we
will see below that the vacuum decay rate in this regime can be
estimated using the stochastic picture.

Presently, let us return to the near-horizon bounces and compute their
tunneling suppression. 
A general expression for it was derived in \cite{Shkerin:2021zbf} and
reads as follows
\begin{equation}
\label{B_Unear_gen}
    B_{U,\text{low-}\l}=\frac{4\pi}{\gc^2}\left( \ln\left[ \frac{4\l^2}{\k
          \Bar{b}_{U}} \right]-4 \right) \;.
\end{equation}
It is not sensitive to the shape of the mode
potential (\ref{Ueff}) in the transition region $|x|\lesssim\l^{-1}$
and, hence, is applicable to our model.
Keeping only the logarithmically enhanced
terms, we obtain 
\begin{equation}
\label{B_Unear_weak}
    B_{U,\text{low-}\l}=\frac{16\pi}{\gc^2}\left( \ln\sqrt{\frac{\l m}{\k}} 
-\frac{\l }{\pi m}\, \H\Big(\frac{q\l}{m}\Big) 
\right) \;.
\end{equation}
For $q=0$, using eq.~(\ref{Green:HL_as1}), we recover the suppression
for the model without dilaton barrier \cite{Shkerin:2021zbf},
\begin{equation}
\label{B_Unear_weak0}
    B_{U,\text{low-}\l}\Big|_{q=0}=\frac{16\pi}{\gc^2}\left( \ln\sqrt{\frac{\l m}{\k}} 
-\frac{8\l }{3\pi m}\right)\;.
\end{equation}
For all values of $q$ corresponding to weak barrier the suppression
monotonically decreases with temperature from the flat-space value
$B_M$ (\cref{B_M}) at $\l\simeq m$ down to $B_M/2$ at $\l=\l_U$, at which
point the near-horizon bounces cease to exist.

\subsection{Stochastic jumps at high temperature}
\label{ssec:U_far}

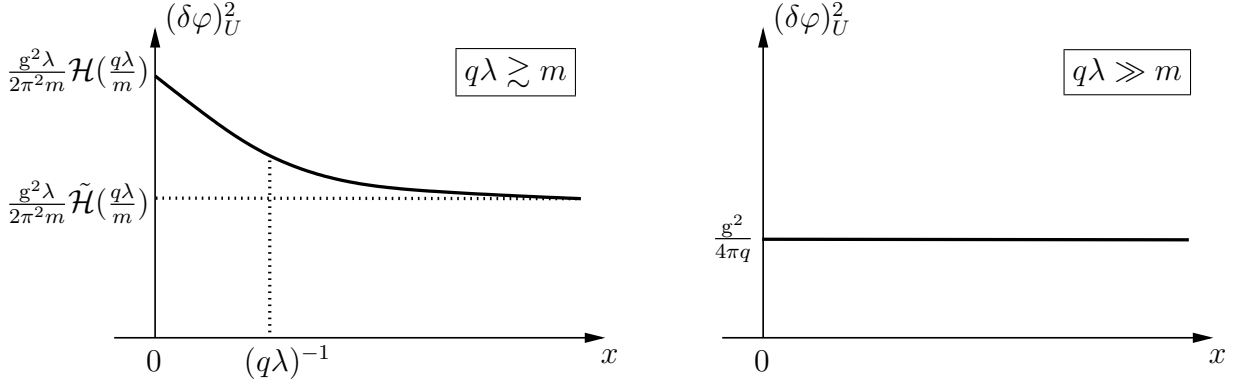
\begin{figure}[t]
    \centering
\begin{picture}(470,140)
  \put(40,20){\includegraphics[width=0.4\linewidth]{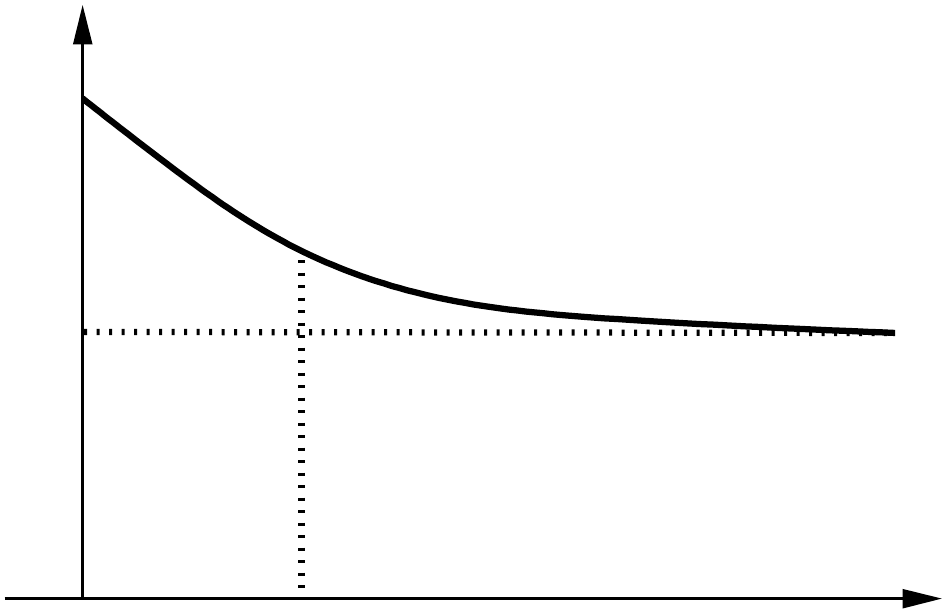}}
\put(0,120){$\frac{{\rm g}^2\l}{2\pi^2m}\H(\frac{q\l}{m})$}
\put(0,70){$\frac{{\rm g}^2\l}{2\pi^2m}\tilde\H(\frac{q\l}{m})$}
\put(60,140){$(\delta\vf)_U^2$}
\put(225,13){$x$}
\put(53,10){$0$}
\put(90,10){$(q\l)^{-1}$}
\put(170,120){$\boxed{q\l\gtrsim m}$}
\put(270,20){\includegraphics[width=0.4\linewidth]{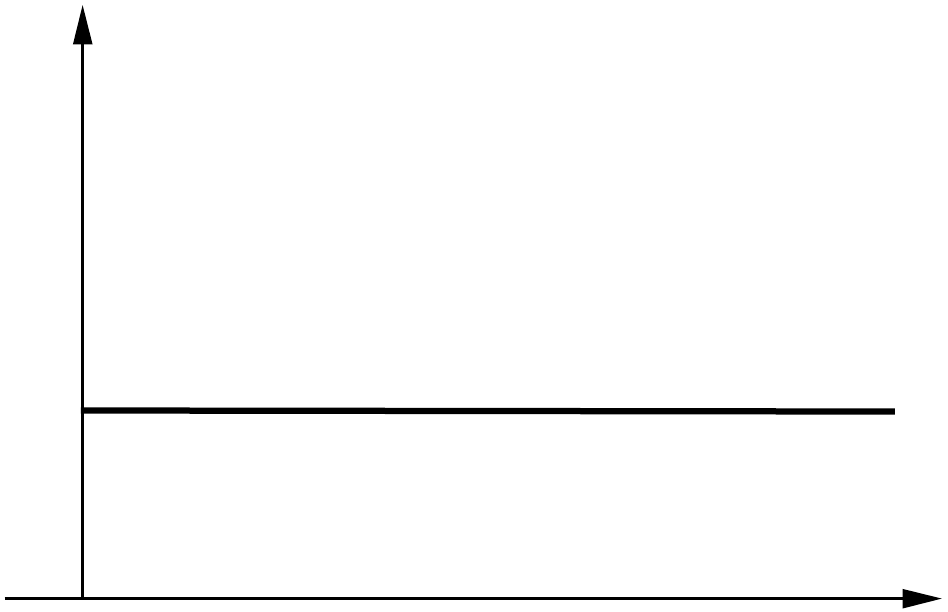}}
\put(267,57){$\frac{\rm g^2}{4\pi q}$}
\put(290,140){$(\delta\vf)_U^2$}
\put(455,13){$x$}
\put(283,10){$0$}
\put(400,120){$\boxed{q\l\gg m}$}
\end{picture}
    \caption{Variance of the field fluctuations in the Unruh vacuum
      outside BH as
    a function of the space coordinate. {\it Left:} moderate BH
    temperature $q\l\gtrsim m$. {\it Right:} limit of very high
    temperature $q\l\gg m$.}
    \label{fig:fluct}
\end{figure}

Following the lessons learned from the Hartle--Hawking case, we expect
that the Unruh vacuum decay at $\l>\l_U$ proceeds via large stochastic
fluctuations kicking the field over the maximum of the scalar
potential $\vf_{\rm max}$. The corresponding decay rate is 
\be
\label{GammaU}
\Gamma_{U,\text{high-}\l}\sim 
\exp\bigg(-\frac{\vf_{\rm max}^2}{2(\delta\vf)_U^2}\bigg)\;. 
\ee
The variance of the fluctuations is estimated from the coincidence
limit of the Green's function,
\be
\label{varU}
(\delta\vf)_{U}^2={\rm g}^2\lim_{t\to 0,\,x'\to
  x}\big[\G_{U}(t,x';0,x)-\G_{F}(t,x';0,x)\big]\;,
\ee 
where for $\G_U$ we use the expressions (\ref{Green:GUright_final}), 
(\ref{Green:GUright_final1}) valid to
the right of the barrier at $x=0$. Focusing on the dominant terms
containing the enhancement factor $\l/m$, we see that the variance
monotonically decreases from the value 
\be
\label{varUnear}
(\delta\vf)_U^2\Big|_{x\to 0}=\frac{{\rm g}^2\l}{2\pi^2 m}
\H\Big(\frac{q\l}{m}\Big)
\ee
in the vicinity of BH down to
\be
\label{varUfar}
(\delta\vf)_U^2\Big|_{x\to +\infty}=\frac{{\rm g}^2\l}{2\pi^2 m}
\tilde\H\Big(\frac{q\l}{m}\Big)
\ee
far away from it. The functions $\H(y)$ and $\tilde\H(y)$ are
defined in eqs.~(\ref{Green:HL}) and (\ref{Green:HR}),
respectively and are plotted in Fig.~\ref{fig:Hs}. 
The characteristic distance from the BH at which the amplitude of the
fluctuations changes
is of order $(q\l)^{-1}$. This is illustrated on the
left plot in Fig.~\ref{fig:fluct}.   

For not-so-large temperature $\l\sim m/q$, the neighborhood of the BH
with large fluctuations is of size $m^{-1}$, sufficient to accommodate
the flat-space sphaleron. At such temperature the
vacuum decay will be dominated by sphaleron transitions in the BH
vicinity. 

At yet higher temperature, $\l\gg m/q$, the size of the region with
enhanced fluctuations shrinks. However, also the variance of the
fluctuations levels out throughout the whole space, because the asymptotics of
the functions $\H(y)$ and $\tilde\H(y)$ at $y\gg 1$ coincide, see
eqs.~(\ref{Green:HL_as2}), (\ref{Green:HR3}). We obtain that in the
high-temperature limit the variance is finite and equals
\be
\label{varUas}
(\delta\vf)_U^2\Big|_{q\l\gg m}=\frac{\rm g^2}{4\pi q}
\ee
irrespective of the position $x$, see Fig.~\ref{fig:fluct}, right
plot.\footnote{The subdominant terms in the Green's function 
(\ref{Green:GUright_final}), (\ref{Green:GUright_final1}) lead to a
slight suppression of $(\delta\vf)^2_U$ at $x=0$ compared to the
asymptotics at infinity. However, the subdominant effects are likely
beyond the validity of the rough
estimates (\ref{GammaU}), (\ref{varU}).} 
This constancy of the variance has clear physical 
interpretation: the flux of particles emitted by the BH and producing
the fluctuations remains constant at arbitrary distance from the BH
due to the two-dimensional nature of the model. Thus, the sphaleron
transitions can now happen with the same probability anywhere in
space. 

Putting everything together, we estimate the rate of the Unruh vacuum
decay at any temperature above $\l_U$ as
\be
\label{GammaU1}
\Gamma_{U,\text{high-}\l}\sim\exp\bigg[-\frac{4\pi^2 m}{{\rm g}^2\l}
\bigg(\H\Big(\frac{q\l}{m}\Big)\bigg)^{-1}
\bigg(\ln\frac{m}{\sqrt\kappa}\bigg)^2\bigg]\;.
\ee
Remarkably, this expression matches the low-temperature suppression
(\ref{B_Unear_weak}) at $\l=\l_U$ up to the first derivative with
respect to $\l$. This supports the stochastic picture of vacuum decay
advocated above. 

\begin{figure}[t]
    \centering
    \includegraphics[width=0.6\linewidth]{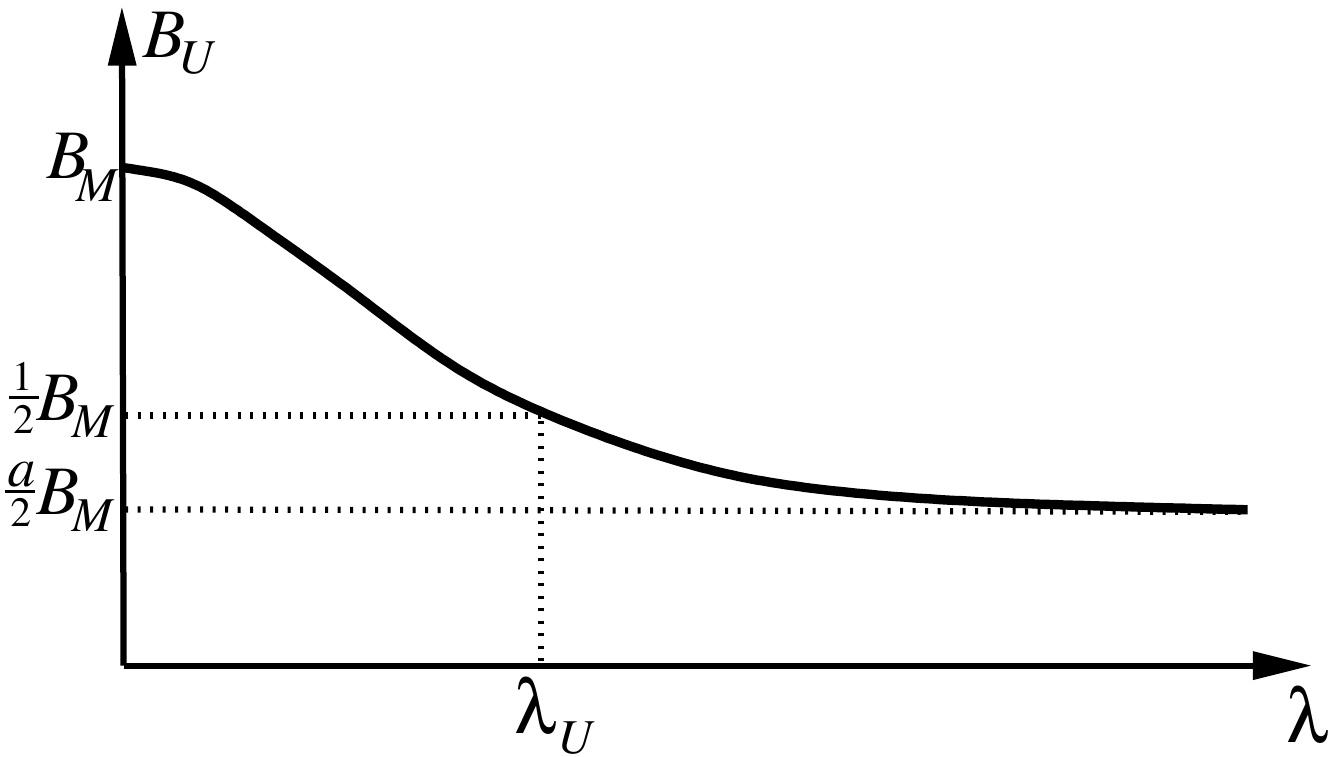}
    \caption{Exponential suppression of the Unruh vacuum decay as a
      function of BH 
      temperature $T_{BH}=\l/(2\pi)$. Decay proceeds via tunneling in
      the near-horizon region at $\l<\l_U$, whereas at $\l>\l_U$ it is
      mediated by stochastic jumps in the vicinity and far away from
      BH. The critical temperature $\l_U$ is given by
      eq.~(\ref{lambdaU}). The suppression approaches a non-zero
      constant in the limit $\l\to\infty$. We consider the case of
      weak dilaton barrier, $0<a<1$.}
    \label{fig:B_U_weak}
\end{figure}

Crucially, the exponential suppression of decay persists even in the
limit of infinite BH temperature,
\be
\label{GammaUinf}
\Gamma_{U,\l\to\infty}\sim\exp\bigg[-\frac{8\pi a}{{\rm g}^2}
\ln\frac{m}{\sqrt\kappa}\bigg]\;.
\ee
This is in striking contrast to the Hartle--Hawking case and is a
direct consequence of the lack of particles in the Unruh flux compared
to the thermal state. We believe this property to be universal and
valid also for BHs in higher dimensions. It hinges on the presence of
the centrifugal barrier reducing the outgoing flux through
non-trivial greybody factors.
We plot our results for the Unruh vacuum decay suppression in
Fig.~\ref{fig:B_U_weak}. 

One point needs to be discussed before closing this section. At very
high BH temperature the Unruh flux is dominated by relativistic modes
with high momenta $k\sim q\l\gg m$ that most efficiently escape
through the barrier. Thus, the correlation length of fluctuations
$l_{\rm corr}\sim k^{-1}$ is much shorter than $m^{-1}$ and further
decreases with temperature. One may ask if this leads to additional
suppression of transitions compared to eq.~(\ref{GammaUinf}), so that
the actual suppression grows with temperature. We now show that this
is not the case, at least in our two-dimensional setup.\footnote{In
  higher dimensions the 
  suppression may actually increase at $\l\to\infty$ due to the
  drop of the particle flux according to the inverse area law at
  finite distance from the BH, see the discussion in sec.~\ref{sec:disc}.}  

To this end, let us look at the spectrum of particles in the Unruh
flux. Far away from the BH particle occupation numbers are given by
the product of the Bose--Einstein distribution and the transmission
coefficient through the barrier,
\be
\label{nkU}
n_k=\frac{1}{\e^{2\pi\omega/\l}-1}\cdot\frac{k}{\omega}|\gamma_\omega|^2\;,
\ee  
where $\omega$ is the particle energy, $\omega=\sqrt{k^2+m^2}$, the
transmission amplitude $\gamma_\omega$ is defined in
eq.~(\ref{Rmodes}), and the factor $k/\omega$ appears due to the different
normalization of plane waves on the left and on the right from the
barrier. Although the relevant particle energies are higher than $m$,
they are still well below $\l$ (recall that $q\ll 1$). Thus, we can
expand the Bose--Einstein factor at $\omega\ll\l$ and use the formula
(\ref{Green:gamma}) for $\gamma_\omega$. We obtain,
\be
\label{nkU1}
n_k=\frac{2\l k}{\pi\big((\omega+k)^2+(q\l)^2\big)}\;.
\ee   
As expected, this describes a broad spectrum of particles centered at
$\omega\approx k\approx q\l/2$.

Let us perform a boost to the rest frame of particles moving with the central
momentum. In terms of the momentum and energy in the new frame the
occupation numbers read
\be
\label{nkU2}
n_{k'}=\frac{m(\omega'+k')}{\pi q\big((\omega'+k')^2+m^2\big)}\;.
\ee
We see that in the boosted frame the Unruh flux represents a
collection of soft particles with momenta $k'\sim m$. Thus, we arrive
at the following physical picture. In the reference frame comoving
with the radiation, the flux consists of soft modes with high
occupation numbers $n_{k'}\sim 1/q\gg 1$. Collisions between these
modes lead to stochastic fluctuations at the scales $k'^{-1}\sim
m^{-1}$ and induce sphaleron transitions in this frame without any
extra exponential suppression on top of eq.~(\ref{GammaUinf}). Note
that from the viewpoint of the original ``laboratory'' frame connected
to the BH, the produced sphalerons are highly boosted.

\section{Conclusions}
\label{sec:disc}

In this paper, we studied a toy model of vacuum decay induced by a
BH. To model the BH background, we used the theory of two-dimensional
dilaton gravity. In this background, we considered the massive scalar
field with the negative Liouville potential. To emulate the
four-dimensional centrifugal barrier for massive scalar modes, we
added the temperature-dependent scalar-dilaton coupling. This coupling
is a new ingredient compared to the model studied in our previous work
\cite{Shkerin:2021zbf}, and it brings us one step closer to a
realistic four-dimensional system. In this model, we first studied Hartle--Hawking vacuum decay, for which our results can also be obtained by applying the standard Euclidean instanton method. In particular, we found that in the high-temperature limit vacuum transitions occur in the asymptotically-flat region and are not suppressed. This is consistent with previous works on BH catalysis \cite{Gregory:2013hja,Burda:2015isa,Burda:2015yfa,Burda:2016mou}.

We then turned to the Unruh vacuum and found that the presence of the barrier changes drastically the exponential suppression of the Unruh vacuum at high temperatures. This is because the barrier reduces the flux of particles emitted by the BH that escape its immediate vicinity. Unlike the Hartle--Hawking case, the suppression of the Unruh vacuum decay does not disappear even in the high-temperature limit; instead, it tends to constant. This is the main result of our study. 

At low temperature the Unruh vacuum decay proceeds via tunneling in
the BH vicinity. We found analytically the corresponding bounce
solution and tunneling rate. At high temperature, the decay regime
changes to stochastic jumps over the sphaleron separating the vacua, and
the decay site shifts to the outer region. Thanks to the features of
our model, we were able to apply a simple stochastic estimate to find
the decay rate in this regime. The stochastic approach can, in
principle, work in more general situations, but will require
full-fledged numerical simulation of the classical field dynamics
\cite{Grigoriev:1988bd,Grigoriev:1989je,Grigoriev:1989ub,Khlebnikov:1998sz}
(see also
\cite{Braden:2018tky,Hertzberg:2019wgx,Hertzberg:2020tqa}). The
implementation of such a simulation would be useful since it would
allow one to go beyond the special exactly-solvable model that was
studied here. 

Note that one important ingredient of the four-dimensional setup ---
the dilution of the Hawking flux as it moves away from the BH 
--- is still not captured by our model. Hence, one
can expect further suppression of the decay probability of the
Unruh vacuum at high BH temperature. Indeed, consider a BH with
$T_{BH}\gg m$ and disregard the effect of the centrifugal barrier,
allowing the occupation numbers of soft modes with $\omega\sim m$,
which are relevant for the decay, to be thermally enhanced 
close to the horizon,
$n_{\rm soft}\big|_{r\sim r_h}\sim T_{BH}/m$. Of course, this is an
overestimate as the greybody factors strongly suppress $n_{\rm
  soft}$. Even in this case, the occupation numbers become small
already at the distance $m^{-1}$, 
$$
n_{\rm soft}\big|_{r\sim m^{-1}}
\sim \frac{T_{BH}}{m}\cdot\bigg(\frac{r_h}{m^{-1}}\bigg)^2\sim
 \frac{m}{T_{BH}}\ll 1\;.
$$
Thus, there are simply not enough modes to generate a
classical field fluctuation that would trigger the decay
\cite{Gorbunov:2017fhq,Johnson_priv}. One can reasonably expect that
the largest catalyzing effect is achieved when the size of the BH is
of order $m^{-1}$. This expectation needs to be confirmed by explicit
calculation, using, e.g., the method of \cite{Shkerin:2021zbf}. We 
leave this for future work.

\section*{Acknowledgments}

We are grateful to Matthew Johnson, Kohei Kamada and Naritaka Oshita
for useful discussions.
The work was in part supported by the Department of Energy Grant
DE-SC0011842 (A.S.). The work was partially supported by the Russian
Foundation for Basic Research grant 20-02-00297 (S.S.). 
The work of S.S. is supported by
the 
Natural Sciences and Engineering Research Council (NSERC) of Canada.
Research at Perimeter Institute is supported in part by the Government
of Canada through the Department of Innovation, Science and Economic
Development Canada and by the Province of Ontario through the Ministry
of Colleges and Universities. 

\appendix

\section{Dilaton black holes}
\label{app:dilaton}

We consider the following action of dilaton gravity in two dimensions
\cite{Callan:1992rs} 
\begin{equation}
    S_{\rm DG}=\int\diff^2x\sqrt{-g}\:\e^{-2\phi}\left(
      R+4(\nabla_\mu\phi)^2+4\l^2 \right) \;. 
\end{equation}
Here $R$ is the scalar curvature and $\l$ is a constant parameter. The
theory admits a one-parameter family of BH solutions with the metric
determined by 
\begin{equation}
    \diff s^2=-\O(r)\diff t^2+\frac{\diff r^2}{\O(r)} \;.
\end{equation}
The function $\O$ and the dilaton field $\phi$ are given by
\begin{equation}
    \O(r)=1-\frac{M}{2\l}\e^{-2\l r} \;, ~~~ \phi=-\l r \;.
\end{equation}
The mass of a BH is $M$ and its horizon radius is
\begin{equation}
    r_h=\frac{1}{2\l}\ln\frac{M}{2\l} \;.
\end{equation}
Introducing the tortoise coordinate
\begin{equation}
    x=\frac{1}{2\l}\ln\left[ \e^{2\l r}-\e^{2\l r_h} \right] - r_h \;,
\end{equation}
and re-expressing $\O$ and $\phi$ as functions of $x$, we obtain
\cref{O,phi}. 

Throughout the paper, we neglect the back-reaction of the tunneling
field $\vf$ on the geometry. This is justified if the gravitational
coupling $\e^{2\phi}$ is 
small compared to the scalar coupling ${\rm g}$ from
eq.~(\ref{action_gen}).
Its maximal value in the BH exterior is reached at the horizon,
\be
\e^{2\phi}\Big|_{r=r_h}=\frac{2\l}{M}\;. 
\ee
This imposes a restriction on the
parameter $M_0$ introduced in \cref{M(T)},
\begin{equation} 
\label{bound_M0}
    M_0^2\gg\frac{\l^2}{\gc^2} \;.
\end{equation}
This can always be satisfied for large enough $M_0$. 
Note that if we allow $q=2Q/M_0^2$ to be of order one, 
the non-minimal coupling of the scalar to the dilaton must be large, 
\begin{equation}
    Q\gg\frac{\l^2}{\gc^2} \;.
\end{equation}
However, this is not a problem since $Q$ enters the action in the
combination $Q\e^{2\phi}$ which is bounded from above by $q\l^2$.

\section{Schwarzschild black hole in four dimensions}

\subsection{Mode potential}
\label{app:Schw}

Let us see how the greybody factors arise in the
theory obtained from four dimensions by a spherical reduction.
Consider a free massive scalar $\chi$ in
four dimensions, 
\begin{equation} 
\label{Schw_action}
	S = \int \diff^4x\sqrt{-g}\left( -\frac{1}{2}g^{\mu\nu}\partial_\mu\chi\partial_\nu\chi
	- \frac{m^2\chi^2}{2} \right) \;,
\end{equation}
and adopt the Schwarzschild metric,
\begin{equation}
\label{Schw_metric}
    \diff s^2=-\O(r)\diff t^2+\frac{\diff r^2}{\O(r)}+r^2\diff\O_2^2 \;,
\end{equation}
where $\diff\O_2^2$ is the line element of a unit 2-sphere and
\begin{equation}
\label{Schw_O}
    \O(r)=1-\frac{r_h}{r} \;.
\end{equation}
Let us make the field redefinition
\begin{equation}
\label{Schw_vf}
    \vf=r\chi
\end{equation}
and introduce the tortoise coordinate
\begin{equation}
\label{Schw_x}
    x=r+r_h\ln\left[\frac{r}{r_h}-1 \right] \;.
\end{equation}
Then, upon restricting to spherically-symmetric configurations,
$\chi=\chi(t,r)$, changing the variables in the action
(\ref{Schw_action}) according to \cref{Schw_vf,Schw_x} and integrating
by parts, we obtain 
\begin{equation}
\label{Schw_action2}
	S = 4\pi\int \diff t \diff x \left( \frac{1}{2}\dot{\vf}^2-\frac{1}{2}\vf'^2 -\frac{1}{2}\left( m^2\O + \frac{\O'}{r} \right)\vf^2 \right) \;,
\end{equation}
where dot (prime) denotes derivative with respect to $t$ ($x$) and
$r=r(x)$ is the inverse of \cref{Schw_x}.  

\begin{figure}[t]
    \centering
    \includegraphics[width=0.6\linewidth]{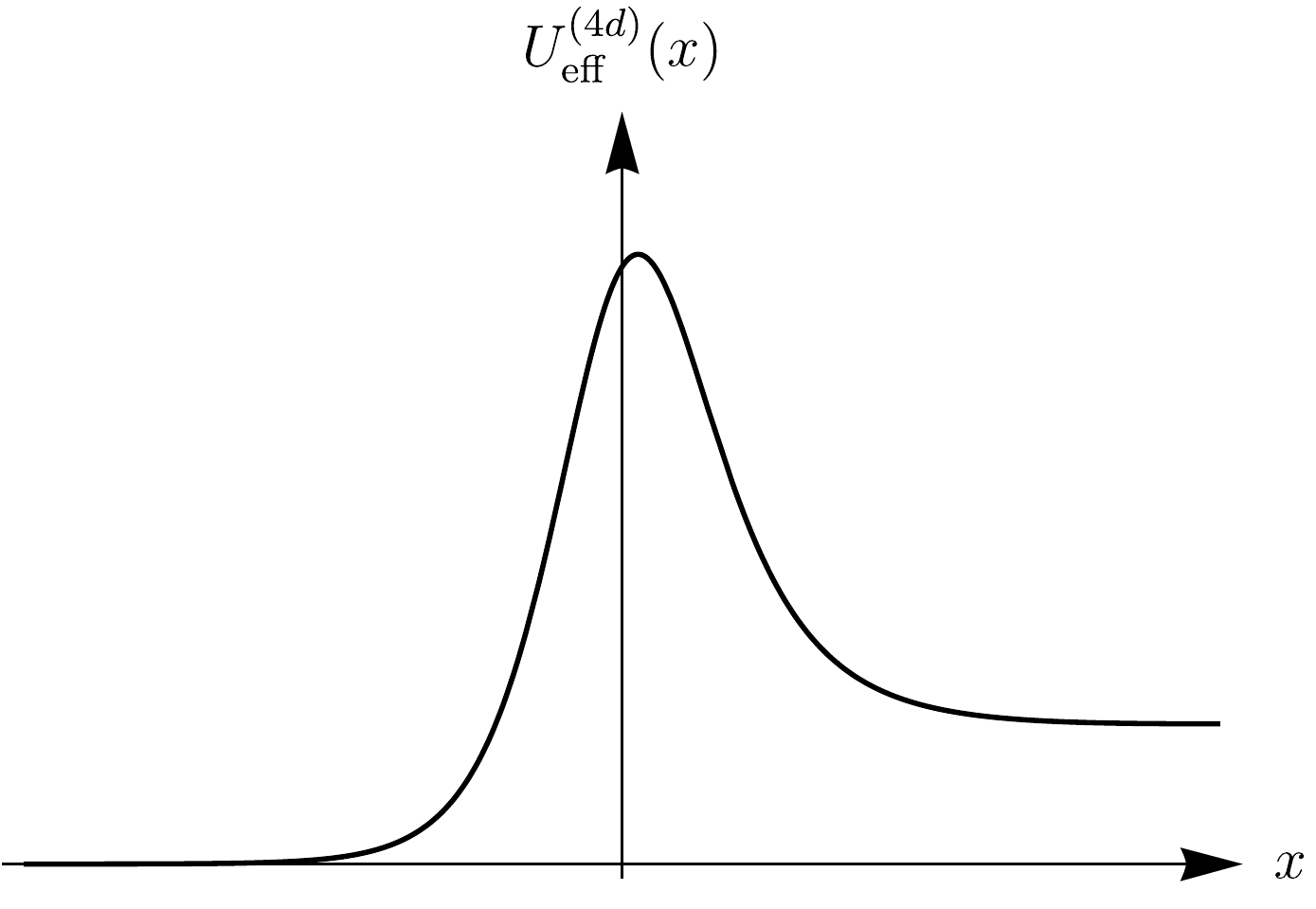}
    \caption{Effective potential for spherically
      symmetric linear massive modes in the four-dimensional
      Schwarzschild geometry.} 
    \label{fig:Schw_pot}
\end{figure}

From \cref{Schw_action2} we read off the effective potential for
spherically-symmetric scalar modes (see Fig.~\ref{fig:Schw_pot} for
illustration), 
\begin{equation}
\label{Schw_Ueff}
    U_{\rm eff}^{(4d)}=m^2\O+\frac{\O'}{r} \;.
\end{equation}
This should be compared with the effective potential in the dilaton
BH (\ref{Ueff_gen}). 
Let us focus on the second term that gives rise to the potential
barrier. In the region $x\sim 0$, where the barrier achieves its
maximum, $r$ is of order the Schwarzschild radius, $r\sim r_h$. The
latter is related to the BH temperature as $r_h=(2\l)^{-1}$. Hence, to
mimic the Schwarzschild greybody factors with two-dimensional dilaton
gravity,     
the coefficient $2Q/M$ in \cref{Ueff_gen} must be $\propto\l$. This is
achieved by imposing the condition (\ref{M(T)}).

Comparing the near-horizon asymptotics of the two potentials, we see
that they agree at $q=2$. However, as explained in the main text, the
physics of vacuum decay is qualitatively
similar already at much lower values of $q$, and the range
(\ref{bound_p}) suffices. The large distance asymptotics of
\cref{Ueff,Schw_Ueff} are different because of the different form of
the function $\O(x)$ and the presence of the
function $r=r(x)$
in the denominator of the second term in \cref{Schw_Ueff}. The latter 
comes
from the area growth in four dimensions. Removing it, i.e., replacing
$r(x)\mapsto r_h$, makes the two potentials completely analogous. In
particular, in the near-horizon region the area factor is not
important.

\subsection{Hartle--Hawking sphaleron}
\label{app:four}

Here we study the Hartle--Hawking sphaleron in the four-dimensional
Schwarzschild background for a scalar theory with inverted quartic
potential. 
Consider
the action 
\begin{equation}
    \label{Sph4:S_gen}
    S=\int\diff^4 x\sqrt{-g}\left(
      -\frac{1}{2}g^{\mu\nu}\partial_\mu\chi\partial_\nu\chi-\frac{1}{2}m^2\chi^2+\frac{{\rm
        g}^2}{4}\chi^4 \right) \;,
\end{equation}
where ${\rm g}^2>0$. 
We redefine the coordinates and the field variable $\chi$ as in
appendix \ref{app:Schw}, further rescale $\vf\mapsto\vf/{\rm g}$, 
and restrict to spherically-symmetric configurations. In this
way, we arrive at 
\begin{equation}
    \label{Sph4_S_sph}
    S=\frac{4\pi}{{\rm g}^2}\int\diff t\diff x \left(
      \frac{1}{2}\dot{\vf}^2-\frac{1}{2}\vf'^2-\frac{1}{2}\left(m^2\O+\frac{\O'}{r}
      \right)\vf^2+\frac{\O}{4r^2}\vf^4\right) \;, 
\end{equation}
where the conformal factor $\O$ is defined in
\cref{Schw_O}, and $r=r(x)$ is the inverse of
\cref{Schw_x}. 
One observes that the structure of this action closely resembles that
of the two-dimensional model (\ref{action_O}) studied in the main
text. The important difference, however, is the explicit coordinate
dependence of the interaction term.

\begin{figure}[t]
    \centering
    \includegraphics[width=0.7\linewidth]{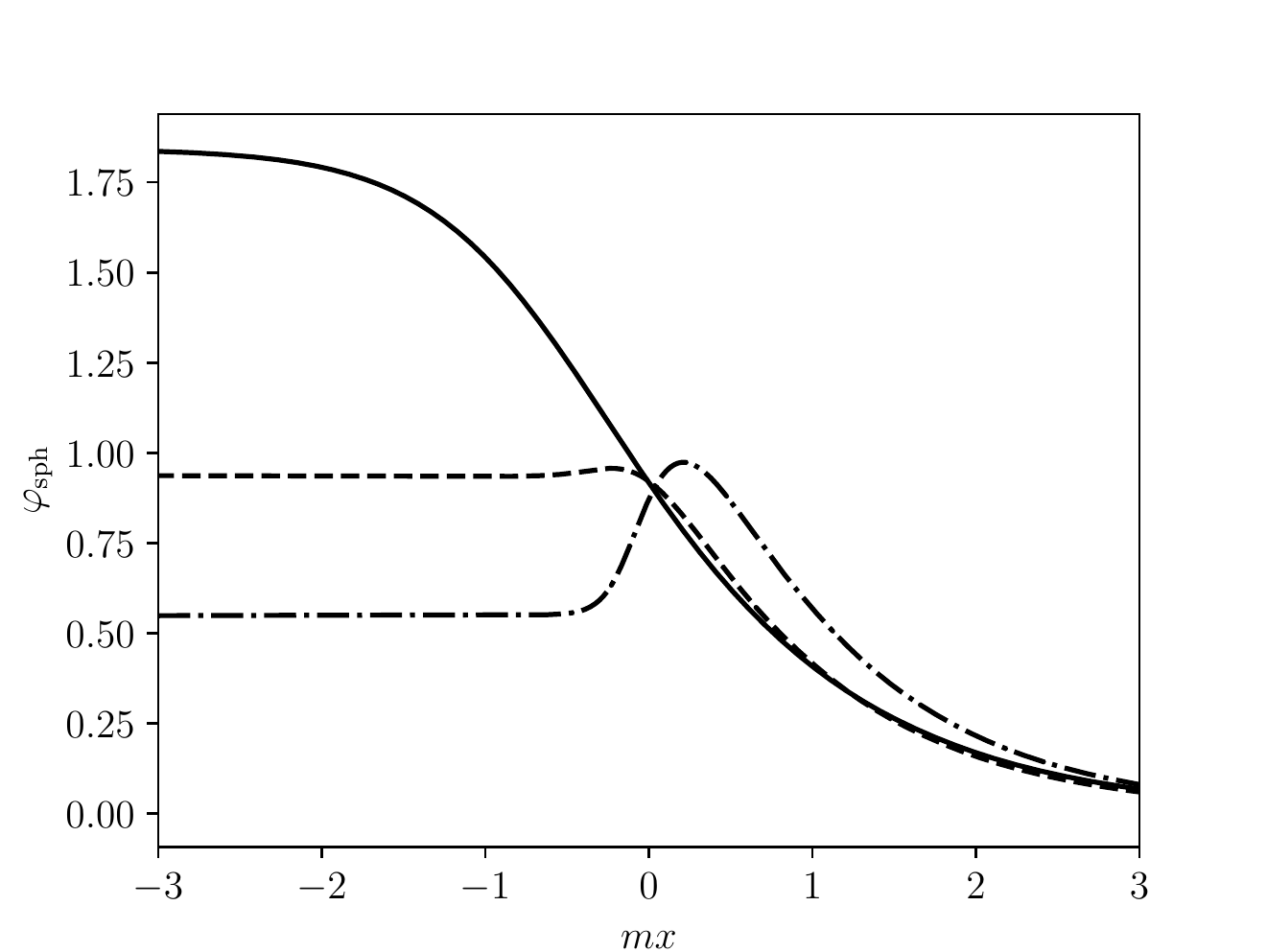}
    \caption{Profiles of the Hartle--Hawking sphaleron in the
      four-dimensional Schwarzschild background at different BH
      temperatures: $2\pi T_{BH}=1.0m$ (solid), $3.5m$ (dashed),
      $7.0m$ (dash-dot).} 
    \label{fig:Sph4_sph}
\end{figure}

The equation of
motion for a static configuration is 
\begin{equation}
\label{Sph4_eom1}
\vf''_{\rm sph}-\left(m^2 \O+\frac{\O' }{r} \right)\vf_{\rm
  sph}+\frac{\O }{r^2}\vf_{\rm sph}^3=0\;. 
\end{equation}
This must be supplemented with the boundary conditions,
\begin{equation}
    \label{Sph4_bc}
\vf_{\rm sph}(x\to-\infty)\to\text{const} \;, ~~~ \vf_{\rm
  sph}(x\to\infty)\to 0 \;. 
\end{equation}
The first condition reflects regularity of the Hartle--Hawking state
at the horizon, and the second is the vacuum boundary condition away
from the BH. Finally, the energy of the sphaleron reads 
\begin{equation}
\label{Sph4_E}
E_{\rm sph}=\frac{4\pi}{{\rm
    g}^2}\int_{-\infty}^{\infty}\diff x\:\left(\frac{1}{2}\vf_{\rm
    sph}'^2+\frac{1}{2}\left(m^2\O+\frac{\O'}{r} \right)\vf_{\rm
    sph}^2-\frac{\O}{4r^2}\vf_{\rm sph}^4\right) \;. 
\end{equation}

We solve \cref{Sph4_eom1,Sph4_bc} numerically. 
We are interested in the behavior of
the sphaleron profile as the Schwarzschild radius $r_h$ (or the BH
temperature $T_{\rm BH}=1/(4\pi r_h)$) varies. The sample plots are
shown in Fig.~\ref{fig:Sph4_sph}. We see that at $r_h m\gtrsim 1$, the
nonlinear core of the solution is localized in the near-horizon
region. On the other hand, at $r_h m\lesssim 1$, the core does not fit
the BH neighborhood. Moreover, the function $\vf_{\rm sph}(x)$ is not a
monotonic function, with the maximum outside the
near-horizon region.\footnote{The sphaleron profile is still monotonic if
  written in terms of the original field variable $\chi$.} 

At small BH temperatures, the Hartle--Hawking sphaleron deviates from
its flat counterpart. This is because the size of the BH $r_h$ exceeds
the characteristic size of the flat sphaleron which is $\propto
m^{-1}$. In the opposite limit the BH is small and, as one can readily
check, the sphaleron is insensitive to the curved geometry and tends
asymptotically to the flat-space solution.  

\begin{figure}[t]
    \centering
    \includegraphics[width=0.6\linewidth]{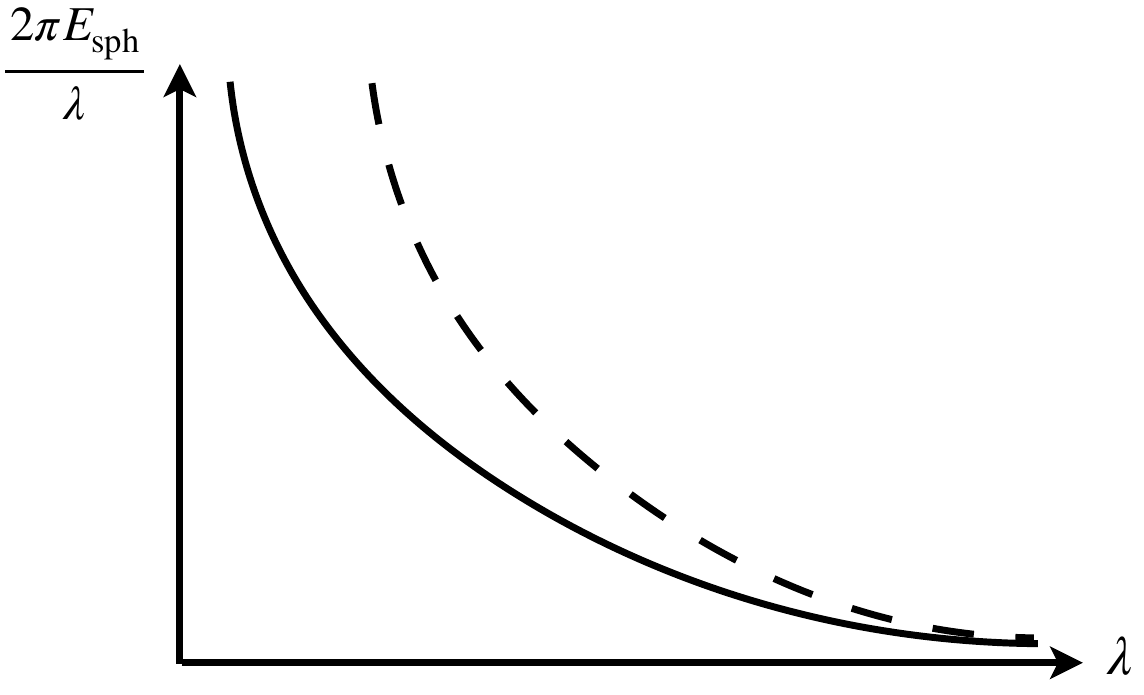}
    \caption{Suppression of the sphaleron transitions in 
the Hartle--Hawking vacuum as function of the BH temperature
$T_{BH}=\l/(2\pi)$ (solid). The sphaleron
      energy $E_{\rm sph}$ is defined in \cref{Sph4_E}.
Dashed line shows the sphaleron suppression in thermal bath at the
same temperature in flat space.} 
    \label{fig:Sph4_B}
\end{figure}

Finally, Fig.~\ref{fig:Sph4_B} shows the Boltzmann suppression factor
$E_{\rm sph}/T_{\rm BH}$ in the BH background vs. the Boltzmann
suppression at the same temperature in flat space. 
We see that the BH transition channel always dominates over the
flat-space channel, although at $r_h m \gtrsim 1$ both are superseded
by the transition via the flat-space periodic bounce \cite{Tetradis:2016vqb}.\footnote{It is
  known that the theory (\ref{Sph4:S_gen}) in flat spacetime does not
  admit a finite-size bounce at zero temperature
  \cite{Affleck:1980mp}. Periodic bounces, however, exist.}
 
Thus, we draw two important conclusions. First, if the BH temperature
is below the scale associated with the size of the sphaleron core
(which is $m^{-1}$ in this case), the
nonlinear part of the sphaleron is localized in the near-horizon
region; at higher BH temperature the sphaleron shifts outside. Second,
in the large 
temperature limit the solution and the associated decay
suppression tend to the ones in flat spacetime. 
Comparing with the two-dimensional model studied in the main text, we see that
this qualitative behavior is reproduced in the regime of weak dilaton barrier,
see sec.~\ref{ssec:HH_far}.

\section{Linear modes and Green's functions}
\label{app:Green}

Here we discuss some further properties of the potential (\ref{Ueff})
and of the linear modes in this potential. We also compute the Green's
functions in the regions of interest. Our aim is to highlight
differences between the model with the scalar-dilaton coupling and the
model without the dilaton barrier studied in \cite{Shkerin:2021zbf}. The
reader is referred to Appendix B of that paper for more details. 

\subsection{Effective potential, modes and scattering coefficients}
\label{app:Green_pot}

Consider the potential (\ref{Ueff}). When $2q\l^2>m^2$, the barrier
generated by the second term exceeds the asymptotics at positive
$x$. The height of the barrier and its position are 
\begin{equation}
\label{Green:vx}
    U_{\rm max}=\frac{(2q\l^2+m^2)^2}{8q\l^2} \;, ~~~ x_{\rm max}=\frac{1}{2\l}\ln\left[ \frac{2q\l^2+m^2}{2q\l^2-m^2} \right] \;, ~~~ 2q\l^2>m^2 \;.
\end{equation}
The width of the region where the potential changes rapidly is
\begin{equation}
\label{Green:width}
    \Delta x\sim \l^{-1} \;.
\end{equation}

As discussed in sec.~\ref{ssec:linear}, the mode equation
(\ref{ModeEq}) with potential (\ref{Ueff}) can be solved exactly in terms of the
hypergeometric functions. Using the general solution, 
we construct a basis of orthogonal 
and delta-function normalizable modes $f_{L,\o}$, $f_{R,\o}$ for
$\o>m$. 
At $x\to\pm\infty$ the modes become plane waves. 
The modes $f_{L,\o}$ are
left-moving at large negative $x$, whereas the modes $f_{R,\o}$
are right-moving at large positive $x$. Using the asymptotics of the
modes $f_{R,\o}$ we determine the reflection and transmission
amplitudes of the potential (\ref{Ueff}), 
\begin{equation}
\label{Rmodes}
f_{R,\o}=\begin{cases}
\e^{i\o x}+\b_\o\,\e^{-i\o x}\;, &x\to-\infty\\
\gamma_\o\,\e^{ikx}\;,  &x\to+\infty
\end{cases}
\end{equation}
where
\begin{equation}
\label{Green:k}
    k=\sqrt{\o^2-m^2} \;, ~~~ \o>m \;.
\end{equation}
These obey the unitarity constraint 
\be
\label{unitar}
|\beta_\o|^2+\frac{k}{\o}|\gamma_\o|^2=1\;.
\ee
The asymptotics of $f_{L,\o}$ are then also fully fixed in terms of
$\beta_\o$ and $\gamma_\o$. 

At $0<\o<m$ only one family of the modes survives, which is
exponentially damped at $x\to+\infty$. This can be obtained from the
modes $f_{R,\o}$ by analytic continuation 
\begin{equation}
\label{Green:vk}
   k\mapsto i\sqrt{m^2-\o^2}\equiv i\vk \;.
\end{equation}
We will continue to denote this family by $f_{R,\o}$, though, of
course, there are no right-moving waves at large $x$ in this
case. The reflection and ``transmission'' amplitudes are still defined
using eq.~(\ref{Rmodes}), but no longer obey the relation
(\ref{unitar}). Instead, we have $|\beta_\omega|=1$.

We only need the expressions for $\b_\o$ and $\g_\o$ at frequencies
much below the temperature $\l$. In this limit, instead of using the
exact hypergeometric mode functions, it is simpler to obtain the
modes by approximating $U_{\rm eff}$ with a superposition of a  
step-function
(corresponding to the first term in \cref{Ueff}) and a 
$\delta$-function (corresponding to the second term). One then finds 
\bseq
\label{Green:betagamma}
\begin{align}
\label{Green:beta}
    & \b_\o=\frac{i(\o-k)+q\l}{i(\o+k)-q\l} \;, ~~~~~ \o\ll\l  \\
\label{Green:gamma}   
    & \g_\o=\frac{2i\o}{i(\o+k)-q\l} \;, ~~~~~ \o\ll\l 
\end{align}
\eseq
where we have assumed $q\ll 1$. The expressions
(\ref{Green:betagamma})
can be analytically continued from $\o>m$ to
$\o<m$ with the replacement (\ref{Green:vk}).

Let us note two properties of the amplitudes $\b_\o$, $\g_\o$. First,
at $q\l\ll\o\ll \l$ one can neglect the term $q\l$ in
eqs.~(\ref{Green:betagamma}) and we return to the case $q=0$. Hence,
$\o\sim q\l$ is the characteristic frequency of the barrier. Second,
in the opposite limit $\o\to 0$ 
the reflection and transmission amplitudes
behave as 
\begin{equation}
\label{Green:ampl2}
    |\b_\o|=1 \;, ~~~ |\g_\o|=\frac{2\o}{m+q\l} \;, ~~~ \o\to 0 \;.
\end{equation}
We observe that the mass in $|\g_\o|$ appears in the 
combination $m+q\l$. The interplay between the two terms in this
combination governs various tunneling regimes, as discussed in the
main text.

\subsection{Green's functions in the asymptotic regions}
\label{app:Green_Green}

Here we summarize the expressions for the Hartle--Hawking and
Unruh Green's functions which we use in the main text, 
postponing their derivation to 
the next subsection. We only need
their form when the two points $x$, $x'$ are placed in the
near-horizon or asymptotically-flat regions, where the mode functions
$f_{L,\o}$, $f_{R,\o}$ are approximated by plane waves. To the former
region we refer as ``left'' and to the latter as ``right''. The two
asymptotics are separated by the region where the potential
(\ref{Ueff}) changes rapidly. The width of this region is given in
\cref{Green:width}. Hence, one can take  
\begin{align}
\label{Green:leftlim}
& x,x'<0\;,~~~~~|x|,|x'|\gg \l^{-1}\qquad\qquad\text{(``left'')}, \\
\label{Green:rightlim}
& x,x'>0\;,~~~~~x,x'\gg \l^{-1}\qquad\qquad~~~\text{(``right'')}.
\end{align}
Furthermore, we need the Green's functions at ``close separation''.
This means that we assume $|x-x'|$, $|t-t'|$ to be sufficiently
small, so that the Green's functions are approximated by their
short-distance asymptotics. The precise conditions are different for
different cases and will be listed below for each case separately.  We will use the superscript ``close'' to indicate that an expression
is valid under this assumption. 

All expressions below are derived under the conditions $m\ll\l$, $q\ll
1$. Note that no relation
between $m$ and $q\l$ is assumed.
Using the time translation invariance
of the BH background, we set $t'=0$.

\paragraph{Hartle--Hawking Green's function (left):}

\begin{equation}
\label{Green:GHH}
\begin{split}
    \left.\G_{HH}\right\vert_{\rm left}= & -\frac{1}{4\pi}\ln\left[4\sh\left(\frac{\l}{2}(x-x'-t)\right) \sh\left(\frac{\l}{2}
(x-x'+t)\right)+i\epsilon\right]\\
& -\frac{\l}{4\pi}(x+x')+\frac{\l}{2\pi (m+q\l)} \;.
\end{split}
\end{equation}
This expression is valid provided that \cref{Green:leftlim} is
fulfilled and $|t|<|x+x'|$. 
No further assumptions about $|x-x'|$ or $|t|$ are needed. 
In the limit $q\to 0$, the Green's
function reduces to the one in the model without the scalar-dilaton
coupling \cite{Shkerin:2021zbf}. We see that the only effect of the
barrier is the replacement $m\mapsto m+q\l$ in the last term in this
expression. 

\paragraph{Hartle--Hawking Green's function (right):}

\begin{equation}
\label{Green:GHH_right}
\begin{split}
    \left.\G_{HH}\right\vert_{\rm right}^{\rm close}= 
& -\frac{1}{4\pi}\ln\left[4\sh\left(\frac{\l}{2}(x-x'-t)\right) \sh\left(\frac{\l}{2}
(x-x'+t)\right)+i\epsilon\right]\\
& +\frac{\l}{4\pi m}+\frac{\l}{4\pi m}\cdot\frac{m-q\l}{m+q\l}\e^{-m(x+x')} \;.
\end{split}
\end{equation}
This expression is valid under \cref{Green:rightlim}, together with
$m|x-x'|,m|t|\ll 1$, $|t|<x+x'$. In the
limit $q\to 0$ it reduces to the expression found in
\cite{Shkerin:2021zbf}. 

\paragraph{Unruh Green's function (left):}

\begin{equation}
\begin{split}
\label{Green:GUleft_final}
    \left.\G_{U}\right\vert_{\rm left}^{\rm close}=- & \frac{1}{4\pi}\ln \left[ 2\sh\left(\frac{\l}{2}(x-x'-t)\right) m
  (x-x'+t)+i\epsilon\right]  -\frac{\l(x+x')}{4\pi} \\
  & +\frac{\l}{2\pi^2m}\H\left(\frac{q\l}{m} \right)
+\frac{1}{8\pi}\H^{(1)}\left(\frac{q\l}{m} \right) \;,
\end{split}
\end{equation}
where 
\bseq
    \begin{align}
       & \H(y)=-\frac{1}{y^2}-\frac{(1+y^2)^2 \arctg{y}}{y^3(y^2-1)}
+\frac{\pi y}{y^2-1} \;, 
\label{Green:HL} \\
       &
       \H^{(1)}(y)=\frac{1}{y^2}-\frac{(1+y^2)^2\ln[1+y^2]}{y^4}+2(\ln
       2-\g_E) \;.  
    \end{align}
\eseq
This formula is valid under the conditions (\ref{Green:leftlim}), 
$|x-x'+t|\ll
\min\{m^{-1},(q\l)^{-1}\}$, and $|t|<|x+x'|$.   

\begin{figure}[t]
    \centering
   \includegraphics[width=0.55\linewidth]{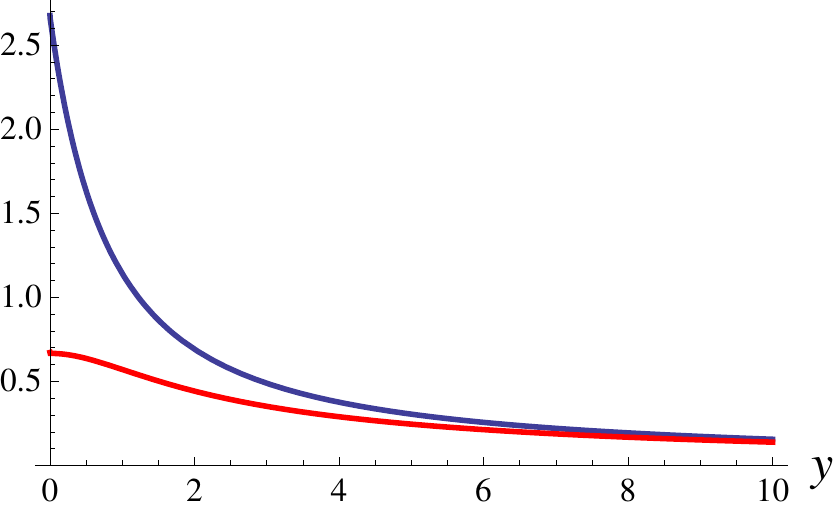}
    \caption{Functions $\H(y)$ (blue) and $\tilde\H(y)$ (red)
      given by eqs.~(\ref{Green:HL}) and (\ref{Green:HR}).}
    \label{fig:Hs}
\end{figure}

The functions $\H$, $\H^{(1)}$ are regular in the limit $y\to 0$:
\begin{equation}
\label{Green:HL_as1}
    \H(0)=\frac{8}{3} \;, ~~~ \H^{(1)}(0)=-\frac{3}{2}+2(\ln 2-\g_E) \;,
\end{equation}
which corresponds to vanishing barrier, and
the function $\left.\G_{U}\right\vert_{\rm left}^{\rm close}$ reduces
to that derived in \cite{Shkerin:2021zbf}. In the opposite limit
$y\to\infty$ the asymptotics of $\H$, $\H^{(1)}$ are 
\begin{equation}
\label{Green:HL_as2}
    \H(y)\approx\frac{\pi}{2y} \;, ~~~ \H^{(1)}(y)\approx -2(\ln y-\ln
    2+\g_E) \;, ~~~ y\to\infty \;. 
\end{equation}
The function $\H(y)$ is monotonic, it is plotted in Fig.~\ref{fig:Hs}.

\paragraph{Unruh Green's function (right):}

\be
\left.\G_{U}\right\vert^{\rm close}_{\rm
  right}=\left.\G_U\right\vert^{\rm close}_{\rm far}+\Delta\G_U\;,
\ee
where
\begin{equation}
    \label{Green:GUright_final}
    \begin{split}
\left.\G_U\right\vert^{\rm close}_{\rm far}=&-\frac{1}{4\pi}
\ln \left[ 2\sh\left(\frac{\l}{2} (x-x'-t)\right) m (x-x'+t)+i\epsilon\right] \\
& +\frac{\l}{2\pi^2m}\tilde\H\left( \frac{q\l}{m}\right)
+\frac{1}{8\pi}\tilde\H^{(1)}\left( \frac{q\l}{m}\right) \;,
    \end{split}
\end{equation}
and 
\be
\label{Green:GUright_final1}
\begin{split}
\Delta\G_U\!=\!\begin{cases}
\frac{\l}{2\pi^2 m}\!\left[\H\!\left(\frac{q\l}{m}\right)
\!-\!\tilde\H\!\left(\frac{q\l}{m}\right)\right]
\!+\!\frac{\l(x+x')}{4\pi}\!\left[\frac{2q\l}{\pi m}
\H\!\left(\frac{q\l}{m}\right)\!-\!1\right]
\!+\!\frac{1}{8\pi}\tilde\H^{(2)}\!\left(\frac{q\l}{m}\right),& x\!+\!x'\ll
\min\{\frac{1}{q\l},\frac{1}{m}\}\\
\frac{1}{2\pi}\ln[m(x+x')]-\frac{\ln
  2-\gamma_E}{2\pi},&\frac{1}{q\l}\ll x\!+\!x'\ll\frac{1}{m}\\
0, &\frac{1}{m}\ll x\!+\!x'
\end{cases}
\end{split}
\ee
Here 
\bseq
\begin{align}
\label{Green:HR}
       & \tilde\H(y)=-\frac{1}{y^2}+\frac{(1+y^2)\arctg{y}}{y^3} \;, \\
       &
       \tilde\H^{(1)}(y)=\frac{1}{y^2}+\frac{(y^4-1)\ln(1+y^2)}{y^4}+2(\ln
       2-\g_E) \;,\\
&\tilde\H^{(2)}(y)=-\frac{2(1+y^2)}{y^2}\ln(1+y^2)\;.
\end{align}
\eseq
Note that the intermediate range in (\ref{Green:GUright_final1})
exists only if $m\ll q\l$.
In all above expressions we assume eq.~(\ref{Green:rightlim}),   
$|x-x'-t|\ll \min\{m^{-1},(q\l)^{-1}\}$, $|x-x'+t|\ll m^{-1}$ and
$|t|<x+x'$. 

In the limit of vanishing barrier, $y\to 0$, the functions
$\tilde\H$, $\tilde\H^{(1)}$, $\tilde\H^{(2)}$ reduce to 
\begin{equation}
\label{Green:HR2}
    \tilde\H(0)=\frac{2}{3} \;, ~~~ 
\tilde\H^{(1)}(0)=\frac{1}{2}+2(\g_E-\ln 2) \;, ~~~ 
\tilde\H^{(2)}(0)=-2\;,
\end{equation}
and the expression for $\left.\G_{U}\right\vert^{\rm close}_{\rm
  right}$ derived in \cite{Shkerin:2021zbf} is reproduced. In the
opposite limit $y\to\infty$ the asymptotics of $\tilde\H$,
$\tilde\H^{(1)}$, $\tilde\H^{(2)}$ are 
\begin{equation}
\label{Green:HR3}
    \tilde\H(y)\approx \frac{\pi}{2y} \;, ~~~ 
\tilde\H^{(1)}(y)\approx 2(\ln y+\ln 2-\g_E) \;, ~~~ 
\tilde\H^{(2)}(y)\approx -4\ln y\;,~~~
y\to\infty \;.
\end{equation}
The function $\tilde\H(y)$ is monotonic and is plotted in Fig.~\ref{fig:Hs}.

\subsection{Calculation of the Green's functions}
\label{app:Green_calc}

\paragraph{Hartle--Hawking Green's function (left).} 

\begin{figure}[t]
    \centering
    \includegraphics[width=0.45\linewidth]{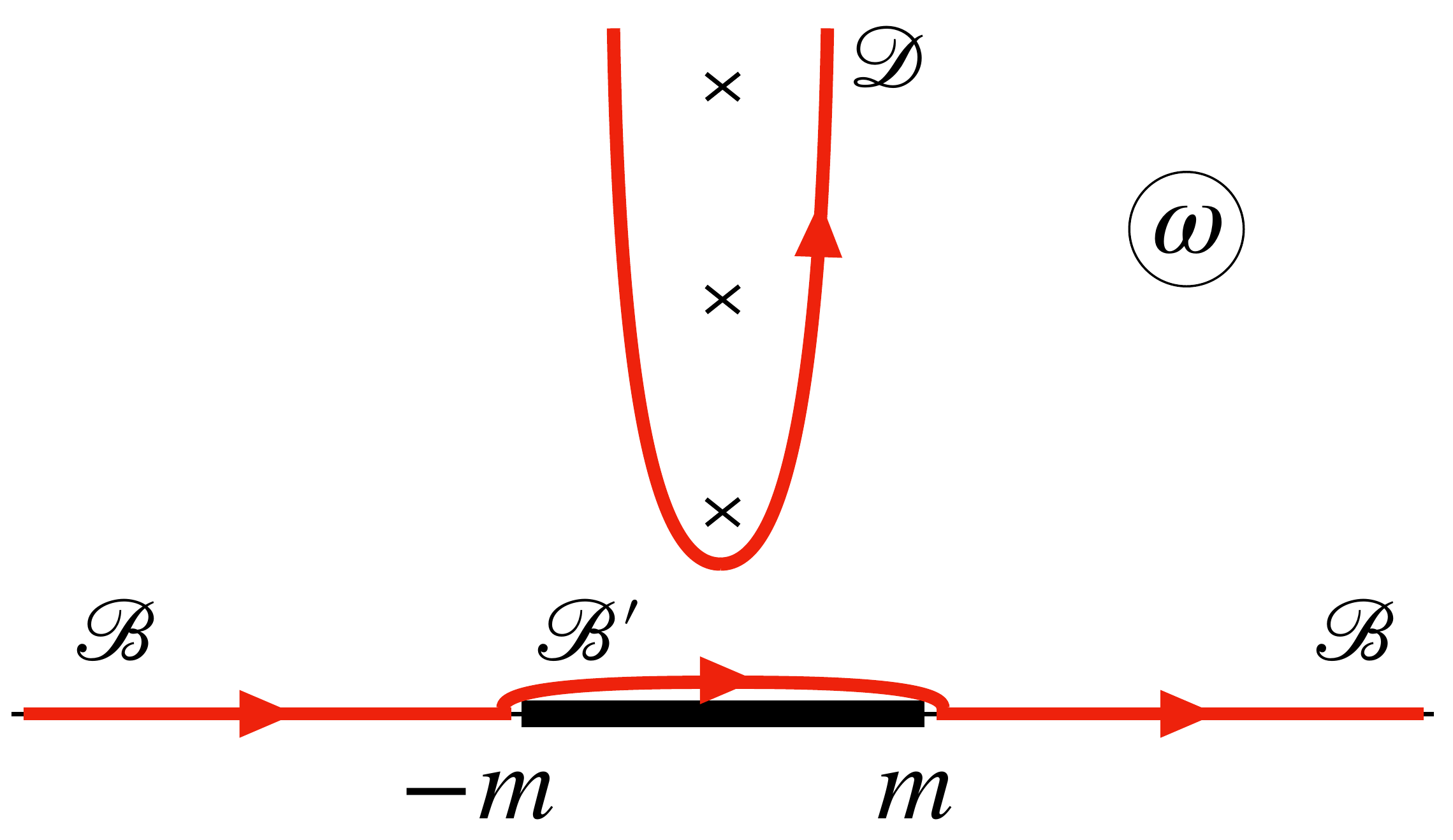}
    \caption{Contours in the $\o$-plane used in the calculation of the
      Hartle--Hawking Green's function and of the Unruh Green's
      function at $q\l<m$.} 
    \label{fig:CHH}
\end{figure}

We start from the general expression for the Green's function in terms
of the reflection amplitude \cite{Shkerin:2021zbf},
\begin{equation}
\label{Green:GHHleft1}
\left.\G_{HH}\right\vert_{\rm left}=\int_0^\infty \frac{\diff\o}{4\pi\o}
\left[2\cos\o(x-x')+\b_\o\e^{-i\o(x+x')}+\b^*_\o\e^{i\o(x+x')}\right]
S(\o)\;,
\end{equation}
where
\begin{equation}
    S(\o)=\frac{\e^{-i\o|t|}}{1-\e^{-\frac{2\pi\o}{\l}}}
+\frac{\e^{i\o|t|}}{\e^{\frac{2\pi\o}{\l}}-1}\;. 
\end{equation}
Let us split the domain of integration in two parts $0\leq\omega< m$
and 
$m\leq\omega<\infty$ and write
\begin{equation}
\label{Green:GHH3}
    \left.\G_{HH}\right\vert_{\rm left}=\G_{HH}^{(1)}+\G_{HH}^{(2)} \;,
\end{equation}
where the integration in the two terms runs over
the first and second domain, respectively. We first
evaluate $\G_{HH}^{(2)}$. Using the identities 
$S(-\o)=-S(\o)$, $\b^*_\o=\b_{-\o}$, it can be brought to the form
\begin{equation}
    \G_{HH}^{(2)}=\int_{\B}\frac{\diff\o}{4\pi\o}\left[ \e^{i\o(x-x')}+\b_\o\e^{-i\o(x+x')} \right]S(\o) \;,
\end{equation}
where $\B=(-\infty,-m]\cup[m,\infty)$, see Fig.~\ref{fig:CHH}. To this
we add and subtract the integral over $\B'$ --- the upper side of
the branch cut at $-m<\o<m$. If $x-x'>|t|$ and $|x+x'|>|t|$, 
the contour $\B\cup\B'$ can be deformed
into the contour $\D$ that encircles the poles of $S(\o)$ at
$\omega=i\l n$, $n=1,2,\ldots$ as shown in Fig.~\ref{fig:CHH}.\footnote{
Note that $\beta_\o$ does not have
singularities in the
upper half-plane due to the absence of bound states in the potential
(\ref{Ueff}).} 
Since $m\ll\l $ and $q\ll 1$, we can
set $\b_\o=0$ when computing the residues over these poles.  We obtain 
\begin{equation}
\label{Green:GHH2}
\begin{split}
        \G_{HH}^{(2)}= & -\frac{1}{4\pi}\ln\left[1-2\e^{-\l(x-x')}\ch{\l t} +\e^{-2\l(x-x')}\right] \\
        & - \int_{-m+i\epsilon}^{m+i\epsilon}\frac{\diff\o}{4\pi\o}\left[ \e^{i\o(x-x')}+\b_\o\e^{-i\o(x+x')} \right] S(\o) \;.
\end{split}
\end{equation}

We now turn to $\G_{HH}^{(1)}$. This can be written as
\be
\label{GHH1}
\G_{HH}^{(1)}=\int_{-m}^{m} \!\!\!\!\!\!\!\!\!\!\!- 
~~~
\frac{\diff\o}{4\pi\o}\left[ \e^{i\o(x-x')}+\b_\o\e^{-i\o(x+x')} \right] S(\o)\;,
\ee 
where the integral is understood in the sense of principal value. We
observe that this almost cancels with the second term in
(\ref{Green:GHH2}), up to a half-residue at the origin. The difference
equals to $-\l x/(2\pi)+\l/(2\pi (m+q\l))$. We combine it with the
first line of eq.~(\ref{Green:GHH2}) 
and analytically continue to $x-x'<|t|$ to obtain \cref{Green:GHH}.

\paragraph{Hartle--Hawking Green's function (right).}
Here the general expression is \cite{Shkerin:2021zbf}
\begin{equation}
\label{Green:GHHright1}
\begin{split}
\left.\G_{HH}\right\vert_{\rm right}=&\int_m^\infty \frac{\diff\o}{4\pi
k}\left[2\cos k(x-x')-\frac{\gamma^*_\o\b_\o}{\gamma_\o}\e^{-ik(x+x')}
-\frac{\gamma_\o\b^*_\o}{\gamma^*_\o}\e^{ik(x+x')}\right]
S(\o)\\
&+\int_0^m\frac{\diff\o}{4\pi\o}|\gamma_\o|^2\e^{-\vk (x+x')}S(\o)
\;,
\end{split}
\end{equation}
where $k$ and $\vk$ are given by eqs.~(\ref{Green:k}), (\ref{Green:vk}).
Note that the last term describes contribution of the non-propagating
modes localized on the BH. We can manipulate the integrals in the
same way as in the previous paragraph. The only new twist are the
restrictions on $|x-x'|$, $|t|$ that must be fulfilled to bring the
result to the final form eq.~(\ref{Green:GHH_right}). We leave the
details of the derivation to the reader.

\paragraph{Unruh Green's function (left).} 

Our starting point is the relation between the Unruh and
Hartle--Hawking Green's functions in the left region \cite{Shkerin:2021zbf},
\begin{equation}
\label{Green:GUleft1}
\left.\G_{U}\right\vert_{\rm left}=\left.\G_{HH}\right\vert_{\rm left}
-\int_m^\infty\dfrac{\diff\o\,k}{2\pi\o^2}\,|\g_\o|^2\,
\dfrac{\cos[\o(x-x'+t)]}{\e^{2\pi\o/\l}-1} \; .
\end{equation}
Let us evaluate the second term which we denote by
$\G_U^{(2)}$. Viewing $k(\o)$ and $|\gamma_\o|^2$ as analytic
functions of $\o$ in the upper half-plane and using $k(-\o)=-k(\o)$,
$|\gamma_{-\o}|^2=|\gamma_\o|^2$ for $\o>m$, we
can bring it to the form
\begin{equation}
\label{Green:GU2}
   \G_{U}^{(2)}= -\int_{\B}\frac{\diff\o\:k}{4\pi\o^2}|\g_\o|^2\frac{\e^{i\o(x-x'+t)}}{1-\e^{-\frac{2\pi\o}{\l}}}+\int_m^\infty\frac{\diff\o\:k}{4\pi\o^2}|\g_\o|^2\e^{i\o(x-x'+t)} \;,
\end{equation}
where $\B=(\infty,-m]\cup [m,\infty)$. Now we add and subtract the
integral over $\B'$---the upper side of the branch cut at $-m<\o<m$. 
Let us assume that $x-x'+t>0$
(we will analytically continue to negative $x-x'+t$ at the end). Then the
contour $\B\cup\B'$ can be deformed into the upper half-plane. This
deformation picks up the thermal poles at $\o=i n\l$, $n=1,2,...$, as
in the case of the Hartle--Hawking Green's function, 
see Fig.~\ref{fig:CHH}. 

\begin{figure}[t]
    \centering
    \includegraphics[width=0.45\linewidth]{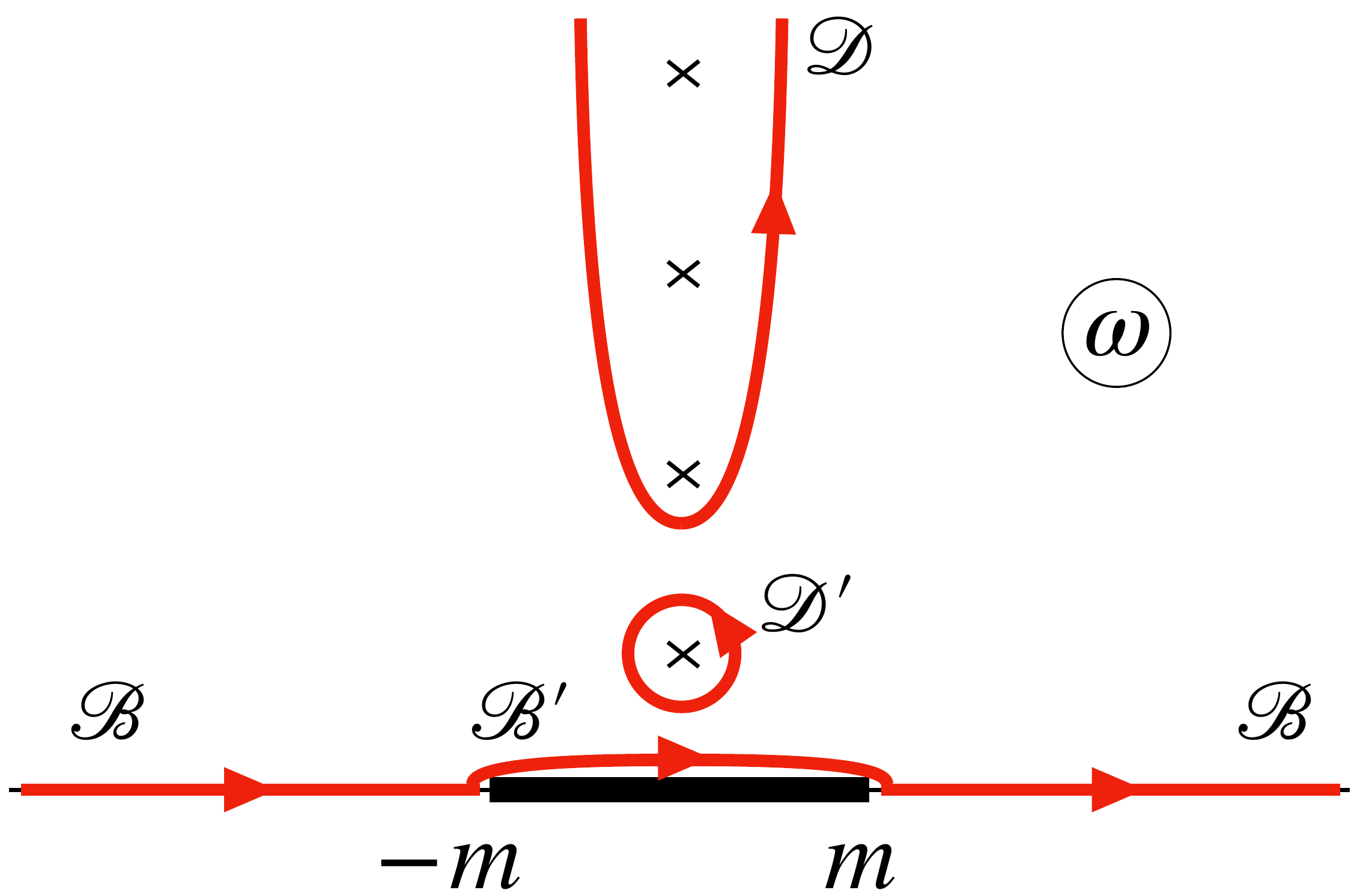}
    \caption{Contours in the $\o$-plane used in the calculation of the
      Unruh Green's function at $q\l>m$. } 
    \label{fig:CU}
\end{figure}

In addition, we should consider the singularities of the transmission
coefficient $|\gamma_\o|^2$.
Using \cref{Green:gamma}, we see that for $q\l<m$ it does 
not have any poles in the upper half-plane. 
On
the other hand, for $q\l>m$ a single pole appears at 
\begin{equation}
\label{Green:res}
    \o_q=\frac{i}{2q\l}((q\l)^2-m^2) \;.
\end{equation}
Since $q\ll 1$, this pole always stays well below the first thermal
pole at $\o=i\l$, see Fig.~\ref{fig:CU}. 

All in all we write,  
\begin{equation}
\label{Green:GU3}
    \G_{U}^{(2)}=\G_{U}^{(21)}+\G_{U}^{(22)}+\G_{U}^{(23)}+\G_{U}^{(24)} \;,
\end{equation}
where the first three terms stand for the integrals along the contours
$\D$, $-\B'$ and $\D'$ shown in Fig.~\ref{fig:CU}, whereas $\G_U^{(24)}$
is just the second term in eq.~(\ref{Green:GU2}).
Let us evaluate these four contributions one by one.

Since $m\ll\l$ and $q\ll 1$, we can set $|\g_\o|^2=1$ in
$\G_{U}^{(21)}$, 
and the result is
\begin{equation}
    \G_{U}^{(21)}=\frac{1}{4\pi}\ln\left[ 1-\e^{-\l(x-x'+t)} \right] \;.
\end{equation}

In $\G_{U}^{(22)}$ we use \cref{Green:gamma} at $\o<m$, assume
$m(x-x'+t)\ll 1$ and expand the integrand to the subleading orders in
$\o/\l$ and $\o(x-x'+t)$ to retain $\mathcal{O}(1)$-contributions. We 
obtain 
\begin{equation}
\begin{split}
    \G_{U}^{(22)}=\frac{i\l}{2\pi^2} \int_{-m+i\epsilon}^{m+i\epsilon}\frac{\diff\o\:\vk}{\o((\o+i\vk)^2+(q\l)^2)} 
     +\left( \frac{i}{2\pi}-\frac{\l(x-x'+t)}{2\pi^2} \right)\int_{-m}^m\frac{\diff\o\:\vk}{(\o+i\vk)^2+(q\l)^2} \;.
\end{split}
\end{equation}
Evaluating the integrals, we arrive at
\begin{equation}
    \begin{split}
        \G_{U}^{(22)}=&-\frac{\l}{4\pi^2m}\left\lbrace \frac{2}{y^2}+\frac{(1+y^2)^2\arctg\left( \frac{2y}{1-y^2} \right)}{y^3(y^2-1)}-\frac{2\pi}{y^2-1}  \right\rbrace \\
        & + \left(-\frac{i}{8}+\frac{\l(x-x'+t)}{8\pi} \right)\left( \theta(1-y)-\frac{1+2y^2}{4y^4}\theta(y-1) \right) \;,
    \end{split}
\end{equation}
where we have denoted
\begin{equation}
\label{Green:y}
    y=\frac{q\l}{m} 
\end{equation}
and $\theta(x)$ is the Heaviside step-function. 

In $\G_{U}^{(23)}$ we assume again that $m(x-x'+t)\ll 1$ and,
moreover, $q\l(x-x'+t)\ll 1$. Evaluation of the residue at $\o=\o_q$
yields
\begin{equation}
    \G_{U}^{(23)}=\left[ -\frac{\l}{4\pi
        m}\frac{(1+y^2)^2}{y^3(y^2-1)} + \left(
        -\frac{i}{8}+\frac{\l(x-x'+t)}{8\pi}
      \right)\frac{(1+y^2)^2}{y^4} \right]\theta(y-1) \;. 
\end{equation}

In $\G_{U}^{(24)}$ we assume that both $m(x-x'+t)\ll 1$ and
$q\l(x-x'+t)\ll 1$ and split the integration domain into $[m,\o_*)$ and
$[\o_*,\infty)$ where $m,q\l\ll \o_*\ll |x-x'+t|^{-1}$. We then
approximate the exponent $\e^{i\o(x-x'+t)}$ by $1$ in the first
sub-integral and approximate the transmission coefficient by $1$ in
the second sub-integral. In this way we get 
\begin{equation}
    \G_{U}^{(24)}=-\frac{1}{4\pi}\ln\left[ m(x-x'+t) \right]+\frac{\ln
      2-\g_E}{4\pi}+\frac{i}{8}+\frac{y^2-(1+y^2)^2\ln[1+y^2]}{8\pi
      y^4} \;. 
\end{equation}

Finally, we combine all four terms together and use the identity
\be
\arctg\left(\frac{2y}{1-y^2}\right)+\pi\theta(y-1)=2\arctg{y}\;.
\ee
Note that, despite the presence of discontinuities at $y=1$ in the individual
terms of \cref{Green:GU3}, the sum is continuous and
smooth at this point. The discontinuity coming with the pole at
$\o=\o_q$ (\cref{Green:res}) that appears in the upper half-plane 
at $y>1$ is
exactly canceled by the discontinuity in the term $\G_{U}^{(22)}$.  

It remains to perform analytic
continuation to $x-x'+t<0$ and add the result to the Hartle--Hawking
Green's function $\left.\G_{HH}\right\vert_{\rm left}$. This leads to
\cref{Green:GUleft_final}.

\paragraph{Unruh Green's function (right).} 
The Unruh Green's function in the right region is related to the
Hartle--Hawking one as follows
\be
\label{GUrightnew}
\begin{split}
\left.\G_{U}\right\vert_{\rm right}=\left.\G_{HH}\right\vert_{\rm right}
&-\int_m^\infty\dfrac{\diff\o}{2\pi k}\,
\dfrac{\cos[k(x-x')+\o t)]+|\beta_\o|^2\cos[k(x-x')-\o
  t)]}{\e^{2\pi\o/\l}-1} \\
&+\int_m^\infty\frac{\diff\o}{2\pi k}
\left[\frac{\gamma^*_\o\beta_\o}{\gamma_\o}\e^{-ik(x+x')}
+\frac{\gamma_\o\beta^*_\o}{\gamma^*_\o}\e^{ik(x+x')}\right]
\dfrac{\cos\o t}{\e^{2\pi\o/\l}-1}\;.
\end{split}
\ee
The second term here, which we denote by $\tilde\G_U^{(2)}$, is computed
following the same steps as in the previous paragraph. We do not
present them in detail and just quote the result,
\be
\label{tildeGU2}
\begin{split}
\tilde\G_U^{(2)}=&\frac{1}{4\pi}\ln\bigg[
\frac{2\sh\frac{\l}{2}(x-x'+t)}{m(x-x'+t)} \bigg]
+\frac{\l}{2\pi^2m}\bigg\{-\frac{1}{y^2}+\frac{1+y^2}{y^3}\arctg{y}
-\frac{\pi}{2}\bigg\} \\
&+\frac{1}{8\pi}\bigg[\frac{1}{y^2}+\frac{y^4-1}{y^4}\ln(1+y^2)
+2(\ln 2-\gamma_E)\bigg]\;,
\end{split}
\ee
where $y$ is defined in \cref{Green:y}. In deriving this expression
we have assumed $m|x-x'+t|\ll 1$, $m|x-x'-t|\ll 1$ and 
$q\l|x-x'-t|\ll 1$. Combining this with the Hartle--Hawking Green's
function (\ref{Green:GHH_right}) far away from the BH, we obtain the
result (\ref{Green:GUright_final}).

The third term in \cref{GUrightnew} is localized in the vicinity of
the BH. Let us denote it by $\tilde\G_U^{(3)}$. Substituting the
expressions (\ref{Green:betagamma}) for the reflection and
transmission amplitudes we obtain,
\be
\tilde\G_U^{(3)}=\int_m^\infty\frac{\diff\o}{2\pi k}\bigg[
\frac{i(\o-k)+q\l}{i(\o+k)+q\l}\e^{-ik(x+x')}+\text{h.c.}\bigg]
\frac{\cos\o t}{\e^{2\pi\o/\l}-1}\;. 
\ee
We observe that the integral converges at
$\omega\lesssim\max\{m,q\l\}\ll \l$. Thus, we can expand the thermal
factor up to the first subleading term to keep track of $\mathcal{O}(1)$
contributions. If we further assume $|t|<x+x'\ll
\min\{m^{-1},(q\l)^{-1}\}$, we can also expand the exponential factor
$\e^{-ik(x+x')}$ and replace $\cos\o t$ with $1$. Then the integrals
are easily taken and we get
\be
\begin{split}
\tilde\G_U^{(3)}=&\frac{\l}{2\pi^2 m}\bigg[\frac{\pi(y^2+1)}{2(y^2-1)}
-\frac{2(1+y^2)}{y(y^2-1)}\arctg{y}\bigg]-\frac{(1+y^2)\ln(1+y^2)}{4\pi y^2}\\
&+\frac{\l(x+x')}{4\pi} \bigg[-\frac{2}{\pi y}+\frac{2y}{y^2-1}
-\frac{2(1+y^2)^2}{\pi y^2(y^2-1)}\arctg{y}\bigg]
\;,~~~~~|t|<x+x'\ll \min\{\tfrac{1}{m},\tfrac{1}{q\l}\}\,.
\end{split}
\ee
If, on the other hand, $x+x'\gg m^{-1}$, the integrand is quickly
oscillating and the integral is damped, so we obtain
\be
\tilde\G_U^{(3)}\approx 0\;,~~~~~~x+x'\gg \tfrac{1}{m}\;.
\ee
The intermediate range $(q\l)^{-1}\ll x+x'\ll m^{-1}$, which exists
only for very high temperatures, requires a careful examination of
various contributions. We leave this exercise to the reader. The
result is 
\be
\tilde\G_U^{(3)}=\frac{\l}{2\pi^2
  m}\bigg(\frac{\pi}{2}-\frac{\pi}{y}\bigg)
+\frac{\l(x+x')}{4\pi}\bigg(-1+\frac{2}{y}\bigg)
+\frac{1}{2\pi} \ln[m(x+x')]-\frac{\ln 2-\gamma_E}{2\pi}\;,
~~~\tfrac{1}{q\l}\ll x+x'\ll \tfrac{1}{m}.
\ee
Combining these expressions with the Hartle--Hawking Green's function
(\ref{Green:GHH_right}) in the BH vicinity, we arrive at
eq.~(\ref{Green:GUright_final1}).

\bibliographystyle{utphys}
\bibliography{refs.bib}

\end{document}